\documentclass[12pt,preprint]{aastex}

\def\deg{\ifmmode^\circ\else$^\circ$\fi}
\def\Fscr{\ifmmode{\mathcal{F}}\else$\mathcal{F}$\fi}
\def\Ascr{\ifmmode{\mathcal{A}}\else$\mathcal{A}$\fi}
\def\spie{Proc. of the SPIE}

\makeatletter

\shorttitle{Atmospheric Transmission}
\shortauthors{C.W. Stubbs et al}
\begin{document}
\title{Towards More Precise Survey Photometry for PanSTARRS and LSST:
Measuring Directly the Optical Transmission Spectrum of the Atmosphere}
\author{Christopher W. Stubbs, F. William High, Matthew R. George,\\
Kimberly L. DeRose, and St{\'e}phane Blondin} 
\affil{Department of Physics \\
and \\
Harvard-Smithsonian Center for Astrophysics \\
Harvard University, Cambridge MA 02138 USA}

\author{John L. Tonry, Kenneth C. Chambers, Benjamin R. Granett}
\affil{Institute for Astronomy, University of Hawaii\\
2680 Woodlawn Drive, Honolulu, HI, 96822 USA}

\author{David L. Burke}
\affil{Kavli Institute for Particle Astrophysics and Cosmology\\
Stanford Linear Accelerator Center\\
Sand Hill Road, Palo Alto, CA, 94025 USA}

\author{R. Chris Smith}
\affil{CTIO/NOAO, 950 North Cherry Ave\\
Tucson AZ, 85719 USA}

\begin{abstract}

Motivated by the recognition that variation in the optical transmission of the atmosphere
is probably the main limitation to the precision of ground-based CCD measurements of celestial fluxes, 
we review the physical processes that attenuate the passage of light
through the Earth's atmosphere. The next generation of astronomical surveys, such as
PanSTARRS and LSST, will greatly benefit from dedicated apparatus to obtain atmospheric 
transmission data that can be associated with each survey image. We review and compare  
various approaches to this measurement problem, including photometry, spectroscopy, and 
LIDAR.  
In conjunction with careful measurements of instrumental throughput, 
atmospheric transmission measurements should allow next-generation imaging 
surveys to produce photometry of unprecedented precision. 
Our primary concerns are the real-time determination of 
aerosol scattering and absorption by water along the line of sight, both of 
which can vary over the course of a night's observations. 

\end{abstract}



\section{INTRODUCTION}

Careful CCD measurements of differential photometry, {\it i.e.} comparing fluxes from similar
stars in the same image, allow comparisons at the millimagnitude level (\cite{Hartman05}, \cite{EH01}). On the 
other hand, with considerable effort the SDSS survey achieved \citep{SDSS07} of 
order 1\% uncertainty in zeropoint uniformity
across the sky in the $g,r,i$ and $z$ bands, and twice that in $u$.  

What accounts for this apparent factor of ten difference 
in our ability to compare the fluxes of celestial objects? The SDSS team attributes \citep{SDSS07} 
the dominant contribution to their zeropoint uncertainty as arising from   ``...unmodelled atmospheric variations... .''  By combining 58 scans across one SDSS equatorial region, \cite{Ivezic07} achieved zeropoint scatter of just under 1\%. This falls short of the $\sqrt{58}~\sim~7$ fold improvement over single-scan calibration that 
one would expect from random errors. This implies that the SDSS photometry is bottoming out in 
some combination of flatfielding residuals and atmospheric variability. 
This fact does not diminish
the importance or the power of the SDSS data set, but future 
survey programs need to identify and overcome the factors
that limited the precision of the SDSS photometry, if they wish to do better. 

When comparing celestial fluxes from objects in a common image, only the {\em angular} variation in 
atmospheric transmission, across the angular separation between objects, can introduce
errors due to differences in atmospheric transmission. On the other hand, comparing fluxes from 
survey images taken at different times and through different airmasses is susceptible to both temporal and line-of-sight
changes in atmospheric transmission. In our view this difference accounts for the discrepancy in 
precision between differential photometry {\em within} one image and establishing a common 
photometric zeropoint {\em across} different survey images. 

This simple fact indicts variations in atmospheric transmission as 
imposing the major limitation to the 
precision of ground-based CCD photometry from all-sky surveys. Disentangling source brightness, 
instrumental response, and variable atmospheric transmission is challenging. Multiple visits
to each field, ideally with different instrument orientations,  
is certainly helpful, along with careful attention to flat-fielding and suppression of instrumental 
artifacts.  

High precision {\em differential} photometry does achieve the Poisson limit. 
Numerous examples of this have been
presented, and we summarize a few salient points from these studies. 
\cite{EH01} performed
aperture photometry on carefully flat-fielded images, and demonstrated that by comparing
fluxes to a robust ensemble average of flux from sources within 0.25 degrees, they 
could achieve Poisson-limited performance. \cite{Hartman05}  used image subtraction 
techniques and demonstrated differential photometry with scatter below 1 mmag. 
Enhanced dynamic range high precision photometry was carried out by \cite{JT05}, 
using an orthogonal transfer CCD. These successes all indicate that flatfielding can be done
at the millimagnitude level, and that Poisson-limited photometry is a worthwhile goal. 
Achieving uniform zeropoints across the surveyed region of the sky 
amounts to making a high precision set of measurements. The survey accuracy, 
{\em i.e.} knowing with certainty the corresponding fluxes in joules/sec/m$^2$, is far
less important (for nearly all astronomical applications) than attaining good precision.

The opportunity for achieving high-precision photometric measurements
in upcoming multiband sky surveys such as PanSTARRS \citep{PS02} and LSST \citep{LSST}
motivates a comprehensive assessment of the limitations of ground-based CCD photometry.
 \cite{ST06} provide a framework for this appraisal, and that paper suggests 
factoring the problem into (i) characterizing the wavelength-dependence of the 
response of the telescope, optics, and instrument, and (ii) determining the optical 
transmission properties of the atmosphere. This paper deals with the issue of
understanding and measuring the optical transmission of the atmosphere.
 
Our eventual goal is to produce the spectrum
of atmospheric optical transmission through which the observation
occurred, for each cataloged object flux from each image in the survey archive. In conjunction with detailed knowledge of instrumental sensitivity versus wavelength
we can then (at least in principle) perform synthetic photometry with trial Spectral Energy Distributions (SEDs) for 
all sources, and compare these spectral integrations with the number of counts 
actually detected. This ``forward modeling'' approach has been recently described for the 
infrared by \cite{Bailey07}. In essence we advocate using the same techniques
that are being applied to correcting for atmospheric absorption in spectrophotometry (e.g. \cite{Bessell99}, \cite{Stritzinger05}) and spectroscopy (e.g.~\cite{Hadrava06}) to broadband imaging data. By reporting the survey fluxes as measured in the ``natural 
system'' of each detector on the focal plane, along with the appropriate instrumental and 
atmospheric transmission profiles, we can construct an atmosphere-corrected
flux for each object of interest using one or more trial SEDs. This 
avoids having to specify color terms for the detectors, which are intrinsically
intermingled with the assumed spectrum of the source of interest. It also 
addresses the problem of second order ``color-airmass'' extinction corrections. 
Users who are satisfied with the more standard treatment are of course
still free to apply a global approximation to instrumental color terms and 
atmospheric extinction corrections. Knowledge of the actual optical transmission 
function for each measurement could also play a role in more precise K corrections for supernovae \citep{ESSENCE}.
 
It is interesting to note that in the era just before the advent of CCD's, photoelectric
photometrists were pushing towards millimagnitude photometry. In this context we
draw particular attention to the remarkable paper by Young and collaborators 
\citep{Young}, that raises many of the issues we address here.  
It would appear that the advent of CCD detectors stalled this initiative. Now that
CCD performance and flatfielding techniques have advanced, it is time to 
revisit the issues that stand between the current state of the art and the 
fundamental limit imposed by Poisson statistics. 

One important difference between our philosophy and that described in \cite{Young}
is that those authors considered a single photometric instrument for both atmospheric characterization 
and celestial flux measurements, whereas we intend to evaluate the optimal choice of 
{\em dedicated} instrumentation for atmospheric characterization. 
Given the choice between using their allocated telescope time to either fully characterize 
atmospheric extinction or to observe their program objects, most astronomers have opted for the latter. 
The SDSS project did 
include a separate ``Photometric Monitoring Telescope" that was equipped with nominally the same
filter set as the 2.5m survey telescope, but filter passband and detector QE differences between the
main survey system and the monitoring telescope gave rise to color terms that 
prevented this approach from reaching its full potential \citep{SDSSpt}.

The Pierre Auger Observatory measures \v{C}erenkov light from high energy cosmic ray showers in the atmosphere. Molecules and aerosols scatter and absorb this signal on the path to the detectors. Proper interpretation of the detected light requires a knowledge of the optical properties of the atmosphere, and this has led the team to establish a sophisticated set of instruments, including LIDAR and photometric monitors to measure the aerosol content and scattering properties of the local atmosphere (\cite{Auger1}, \cite{Auger2}). 

The photometric calibration plans for PanSTARRS \citep{MagnierPS} and LSST \citep{Burke06}
both include apparatus dedicated to the determination of atmospheric extinction, to allow
the wide field survey imagers to focus on science images of the sky without needing to 
allocate time to atmospheric characterization. 

The question then becomes: What is the best method to determine atmospheric extinction, 
using dedicated apparatus that runs in conjunction with a broadband all-sky survey, 
and what cadence of observations is needed to capture the temporal and directional
variations in extinction?

\subsection{Some Formalism} 
 
The total photon flux ($S(t)$, photons per sec)  we detect from some celestial source is an integral over wavelength,

\begin{equation}
S(t) =  \int \Fscr(\lambda, t)
   \times A \times T(\lambda, \hat{z}, t)\times  I(\lambda, t) ~ d\lambda,
\label{eq:psignal}
\end{equation}

\noindent
where the $\Fscr$ is the spectral photon distributions (SPDs, evaluated
above the atmosphere) of the sky and all sources present,
$I(\lambda,t)$ is the dimensionless instrumental transmission, 
including optics, filter, and detector, $A$ is the effective collecting area of the system, 
and   
$T(\lambda, \hat{z}, t)$ is the wavelength dependent atmospheric transmission along the direction $\hat{z}$ at the
time $t$ of the measurement, 
averaged over the exposure time.  (We adopt the convention here that the units of $\Fscr$
are photons/nm/sec/m$^2$, derived from the conventional spectral
energy distribution as $\Fscr = F_\lambda\;\lambda/hc$.)  

We have described elsewhere \citep{CTIOlaser} a program of measuring explicitly $I(\lambda,t)$, the
wavelength dependence of the instrumental response. Our focus in this paper is the 
determination of $T(\lambda, \hat{z}, t)$, the time-dependent and 
line-of-sight dependent optical transmission of the atmosphere.

The claim that observing conditions are ``photometric'' during a night amounts to assuming that 
$T(\lambda, \hat{z},t)=T(\lambda, \theta_{zenith})$, {\it i.e.~}time-independent and axisymmetric,  
depending only on zenith angle.  Furthermore, the typical corrections for 
atmospheric extinction assume a simple dependence in which all magnitudes
are corrected by a band-specific extinction coefficient $k$ that is multiplied by the secant of the 
zenith angle. The extinction coefficient for each passband is taken as a universal 
number, independent of the source's spectrum. As shown below, these standard photometric approximations will fail to produce 
the accuracy we propose to attain. Furthermore, the regression of all magnitudes to 
zero airmass (above the atmosphere) introduces a strong leverage on any uncertainty in the extinction coefficients,  as we are extrapolating into a regime that is not directly observable from 
the ground.  

\cite{Wade88} present a useful formulation of atmospheric 
transmission, which we will generalize to make clear the distinct
processes at work, and any potential time dependence.   
At any wavelength the optical transmission of the atmosphere can be represented as

\begin{equation}
T(\lambda,\hat{z},t)=exp\left [~\sum_i(-\tau_i(\lambda,t,\hat{z})(\chi^{\alpha_i}(\hat{z}))\right ],
\end{equation}

\noindent
where $\tau_i$ is the optical depth (at one airmass) for each attenuation process, $\chi$ is the airmass
along the line of sight with unit vector $\hat{z}$, and $\alpha_i$ is the airmass dependence
of the attenuation process.  For spectral regions where the attenuation processes are unsaturated 
we would expect $\alpha_{unsat}=1$ whereas for saturated lines we would expect the 
absorption from the wings of the line to scale with $\alpha_{sat}=0.5$. 
Atmospheric water has numerous narrow molecular absorption features; 
some are saturated and some are not. \cite{Wade88} suggest that this 
produces an effective $\alpha_{water}=0.6$.  We should expect similar 
behavior from other complex molecular absorption band structures. 
 
The direct determination of each component of atmospheric attenuation, ideally co-boresighted and 
simultaneous with the survey imager, would help address our currently inadequate correction for  
the atmosphere. Our goal in this paper is to explore alternative approaches to this 
measurement problem, and to suggest measurements that would help illuminate a 
shrewd choice of technology.    

Measurements of the optical transmission properties of the atmosphere can exploit either
natural celestial sources to backlight the atmosphere, or manmade illumination which is 
observed through backscatter. (\cite{Albert06} have undertaken a program to use
ground-based observations of artificial sources on satellites, but we will not explore
that option in this paper.) 

The resulting atmospheric 
transmission measurements can then be used in isolation, 
or in conjunction with detailed computer models such as MODTRAN \citep{2001SPIE.4381..455A}. We suggest that 
it makes sense to fully exploit the comprehensive knowledge of atmospheric physics that
has been incorporated into these sophisticated codes. 

In the sections that follow we review the physics of optical transmission through the 
atmosphere, and then we discuss in turn merits of spectroscopic, photometric and LIDAR
measurements for characterizing extinction. 
We then describe how the combination of atmospheric transmission 
functions and computer modeling could be exploited in a forward modeling approach to astronomical photometry, and 
we close with a suggested set of next steps. 

\section{ATTENUATION BY THE ATMOSPHERE AT OPTICAL WAVELENGTHS, AND ITS
VARIABILITY}

\subsection{The Final Four for a Photon: Rayleigh Scattering, Molecular Absorption, Aerosols and Clouds}

As light propagates to us from celestial sources it experiences
numerous opportunities to interact with material. Extinction at the source, 
through the Galaxy, and in other intervening regions can all play a role in 
distorting the spectrum. From the standpoint of this paper, these astrophysical
processes can be either sources of information or a nuisance, and we will
not address them here. Our concern is the astrophysically uninteresting 
attenuation from the final four hurdles faced by an astronomical photon:
scattering and absorption from atoms and molecules in the atmosphere, 
scattering from suspended aerosols, and extinction due to clouds.  

\begin{figure}
\plotone{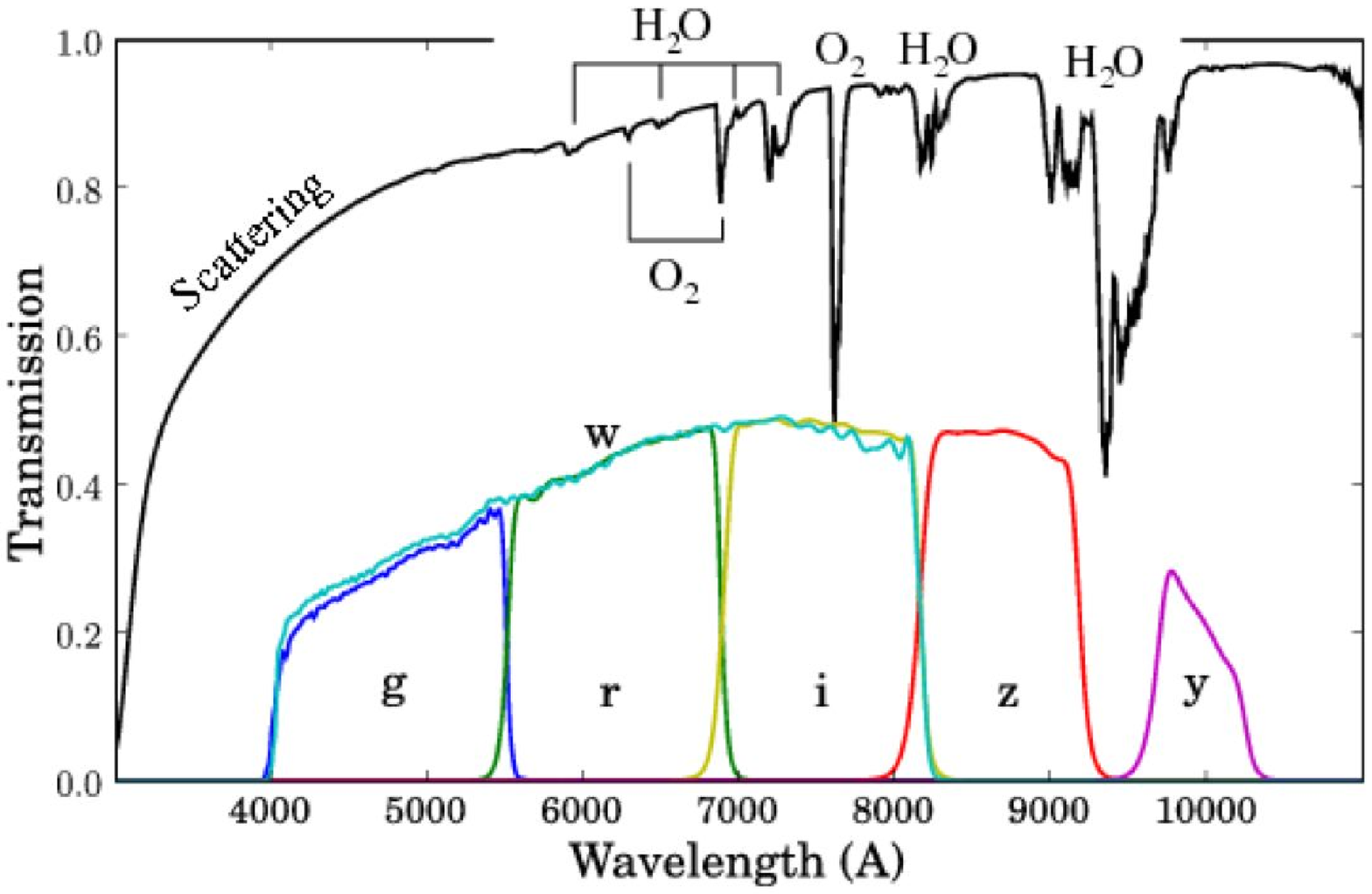}

\caption{Photon Transmission through One Airmass, with PanSTARRS Instrumental 
Sensitivity. 
This plot shows the 
different contributions to attenuation of light in passing through the atmosphere, 
along with filter bands times the expected detector QE, for PanSTARRS. The LSST filter set is similar, but also 
includes a $u$ band. The atmospheric transmission was computed with MODTRAN for one airmass 
at an elevation of 10,000 ft (305m), with an initial spectral resolution of 1 cm$^{-1}$, boxcar smoothed to
1 nm.} 
\label{fig:photon_transmission}
\end{figure}

Figure ~\ref{fig:photon_transmission} shows the expected attenuation of flux due to the
atmosphere, alongside the filter passbands for the PanSTARRS survey, and
illustrates where the different components of extinction will afflict our measurements. 

Table~\ref{tab:smooth} provides an estimate of the integrated attenuation we expect 
within each survey passband, from the different components of 
atmospheric attenuation. This calculation presumes a source that has a 
photon spectrum $\Fscr(\lambda) \propto \lambda^{0}$, which for our spectral region
of interest approximately corresponds to a blackbody of 4500~K. We integrated
the different components of $T(\lambda)$ separately in order to 
gain some intuition about the relative importance of scattering vs.~absorption.

\begin{table}[htdp]
\caption{Estimated transmission at one airmass from the smoothly varying 
components (Rayleigh scattering and aerosols) and absorption from molecular lines, 
for the PanSTARRS system. 
 Note that the airmass dependence 
$\alpha$ of these can be quite different.
We have included the anticipated effects of 
two Al reflections, 
of filter transmission, and of detector QE in the system's response 
functions. These numbers presume
a source with a constant photon flux per nm. The effects of molecular 
absorption and aerosols are comparable for $i, z$ and $y$.}
\begin{center}
\begin{tabular}{cccc}
\hline
Band & T$_{smooth}$ & T$_{lines}$ & Total \\
\hline
PanSTARRS $g$ & 0.815 & 1.000 & 0.815 \\
PanSTARRS $r$ &0.894 & 0.996 & 0.890 \\
PanSTARRS $i$ & 0.949 & 0.961 & 0.912\\
PanSTARRS $z$ & 0.964 &0.970  & 0.935 \\
PanSTARRS $y$ & 0.961 & 0.947  & 0.910\\
\hline
\end{tabular}
\end{center}
\label{tab:smooth}
\end{table}%

The next step is to assess how different source spectra 
behave as they propagate through the atmosphere and the 
instrumental response function. 
Since it is awkward to obtain high precision spectrophotometric data 
in the attenuated spectral regions of interest we used 
theoretical models of stellar spectra for computing the effects of the atmosphere. 
We used the theoretical spectra from \cite{Kurucz} to compute the 
flux attenuation due to the atmosphere by performing synthetic 
photometry across the different $I(\lambda)$ system response 
functions we expect for PanSTARRS. We obtained Kurucz's modelled
stellar spectra from 
{\tt http://www.stsci.edu/hst/observatory/cdbs/astronomical\_catalogs.html}.
These STSCI stellar atlas data are 
$F_{\lambda}(\lambda)$ tabulations, which we converted into photon 
spectral distributions $\Fscr(\lambda)=F_{\lambda} \times \lambda$, with an 
arbitrary common multiplicative normalization.

Using a spectral resolution of $\Delta \lambda =~$0.1~nm we integrated
the source spectra shown in Table~\ref{tab:extinct_vs_spect} through the transmission function
that corresponds to $\chi=~$1 airmass.  
 The {\it error} that would be incurred by making the usual assumption that the 
extinction (in magnitudes) scales linearly in airmass for all attenuation processes 
is roughly 0.01, 0.02 and 0.03 magnitudes per airmass in the $r,i, $ and $z$
bands, respectively. 

\begin{table}[htdp]
\caption{Calculated extinction coefficients for PanSTARRS 
passbands, for different astronomical objects. The first column lists 
the template spectrum used, the subsequent columns list the computed
extinction coefficient, in magnitudes per airmass. The interplay
between color and extinction is most pronounced in the bluer
bands. The difference in spectral weighting accounts for the 
difference between these values and those of the previous
Table.}
\begin{center}
\begin{tabular}{lccccc}
\hline
Source & $k_g$ & $k_r$ & $k_i$ & $k_z$ & $k_y $\\
\hline
O5V & 0.23 & 0.13 & 0.08 & 0.06 & 0.07 \\
B5V  & 0.23 &0.13 & 0.08 & 0.06 & 0.07 \\
A3V  & 0.23 & 0.13 & 0.08 & 0.06 & 0.07\\
F5V  & 0.22 & 0.12 &0.08  & 0.06 & 0.07 \\
G3V  &0.21 & 0.12 & 0.08  & 0.06 & 0.07\\
K4V & 0.20 & 0.12 & 0.08 & 0.06 & 0.07 \\
M2V  & 0.20 & 0.12 & 0.08 & 0.06 & 0.07\\
\hline
\end{tabular}
\end{center}
\label{tab:extinct_vs_spect}
\end{table}%

We drew two conclusions from this exercise: 1) we cannot ignore the interplay
between extinction and the source spectrum, and 2) we need to properly account
for the different airmass scaling for line absorption vs. scattering.

A number of studies have explored the stability of extinction at astronomical sites, 
including \cite{KrisciunasMK}, \cite{Riemann92}, \cite{Frogel}, \cite{SanPedro}, 
\cite{DeVauc73} and \cite{SDSSpt}.
There are numerous reasons to expect variability in atmospheric extinction. 
We should expect a variation in aerosol and water content 
with the direction of prevailing winds and with local meterology.
The solar ``weather'' influences the ozone fraction in the upper atmosphere.  
Some of these dependencies are nontrivial. One such example is the apparent
dependence (\cite{Pakstiene}, \cite{Riemann92}, \cite{Roosen77}) 
of extinction on absolute humidity, even in bluer spectral regions that
are well away from H$_2$O line absorption, allegedly due to water absorption by 
aerosols changing their sizes and hence their scattering properties. 
The dependence of extinction on meteorology, and over time, has been explored by \cite{Riemann92}, 
\cite{Frogel}  and \cite{Pakstiene}. 

Table~\ref{tab:ctio} shows a summary of weather statistics for a 4 year period at CTIO. We will take these
values as being representative of weather variability at good astronomical sites around the world. 
The data were obtained from the online archive at\\ 
{\tt http://www.soartelescope.org/release/02about/eng$\_$about/weather/main$\_$weather.html}

\begin{table}[htdp]
\caption{Summary of Weather at CTIO, Jan 1 2001 to Jan 2 2005. The table lists 
percentiles for the distribution of the relevant meteorological parameter. We make the conservative
assumption that the time actually spent observing follows this overall pattern. }
\begin{center}
\begin{tabular}{cccccc}
\hline
~&10th&25th&50th&75th&90th\\
\hline
Temperature (C)& 6.7& 11.9& 15.2& 17.4& 18.9\\
Pressure (mbar)& 781&782&784&785&786\\
Rel Humidity (\%)&8&13&23&38&55\\
Wind Direction (Degrees) & 32& 60& 107& 238& 313\\
Wind Speed (m/s)&  0.675& 1.62& 3.195& 5.85& 8.595\\
\hline

\end{tabular}
\end{center}
\label{tab:ctio}
\end{table}%

\subsection{Molecular  scattering}

Elastic scattering \citep{Rayleigh} from atoms and molecules in the air
has a cross section that varies as $\sigma_{Ray}(\lambda) \sim \lambda^{-4}$, 
with a slight correction that arises from wavelength dependence of the index of refraction.
\cite{Hansen} give the optical depth (at zenith) for Rayleigh scattering to be 
$$
\tau_{Rayleigh} =  0.008569 \lambda^{-4}\left (1+0.0113 \lambda^{-2} + 0.00013 \lambda ^ {-4}\right ) \times \left [\frac{P}{1013.25~mb} \right],
$$
\noindent
where $\lambda$ is in microns.  This process dominates over inelastic scattering, such that the 
inelastic contribution to attenuation is negligible.

The attenuation due to Rayleigh scattering thus depends on 
the pressure of the atmosphere along the line of sight. 
Barometric pressure varies with a typical timescale
of days, and will fluctuate as high and low
pressure systems pass over an observatory. This changes the effective airmass 
along a fixed line of sight, in proportion to the pressure fluctuation.  
A measurement taken at CTIO at an airmass $\chi$ under a pressure that is 
$\delta P$ different from the nominal $P_o=$784 nm will 
suffer a change in transmission of $\delta T_{Rayleigh} = 6 \times 10^{-3}~\chi~\left(\frac{\delta P}{P_o}\right)\left(\frac{1~\mu m}{\lambda}\right)^4$.
A variation in attenuation of less than one millimagnitude 
therefore requires $\delta T_{Rayleigh} < 0.001$ or $\chi~\left(\frac{\delta P}{P_o}\right)\left(\frac{1~\mu m}{\lambda}\right)^4~<~0.17$. Using the 90$^{th}$ percentile 
pressure excursion for CTIO, an {\it uncorrected} Rayleigh extinction perturbation under 1 millimagnitude at  $\lambda=$500 nm, in the $g$ band,  corresponds to 
restricting $\chi < $ 2.6 airmasses during a large pressure excursion.

It is however very tractable to calculate how changes in local barometric pressure 
would introduce  
an airmass-dependent shift in the Rayleigh transmission. 
Millimagnitude photometry will
require a pressure-dependent correction to attenuation from Rayleigh scattering 
only for very blue bands observed at airmasses around $\chi\sim3$. We are also fortunate that conditions
with very low barometric pressure are often accompanied by weather that precludes opening the dome. 
The Rayleigh scattering should 
be axisymmetric about the zenith with a time dependence driven only by pressure
variations. 

The characteristic spectral dependence and spatial symmetry of extinction by molecular scattering can be combined with independent measurements of barometric pressure and airmass to provide precise determination of extinction due this process, subject
to observational confirmation. 

\subsection{Aerosols}

Scattering from aerosols and particulates in the atmosphere is in the awkward 
``Mie'' regime where the particulate size is comparable to the wavelength. This 
gives rise to a cross section $\sigma_{Mie}(\lambda)  \sim \lambda^{k}$ where $k$ 
depends on the size and shape distributions of the scattering particles. Since aerosol scattering is in the optically thin regime for all 
wavelengths, we expect an airmass exponent of $\alpha_{aerosol}=1$. \cite{Sinyuk07}
claim that most aerosols reside within $\sim $ 4 km of the Earth's surface.

There is ample evidence for volcanic eruptions producing long term change in the optical 
transmission of the atmosphere. These global events have 
more local counterparts, due to 
changes in wind direction, to regional forest fires, to lofted marine salts, etc. This implies 
that the transmission spectrum $T_{aerosol}(\lambda, \hat{z},t)$ will have both time dependence 
and azimuth dependence. 

\cite{Holben99} present data for both atmospheric water content and the optical depth due to aerosols, obtained from the AERONET system, 
at Mauna Loa. We strongly suspect this is a good proxy for the aerosol
characteristics we expect on Haleakala, where PanSTARRS-1 is situated.  
The AERONET instrument measures the scattering of solar
radiation as a function of angle away from the sun, in multiple bands. 
Although the observations in the different bands are not strictly simultaneous, 
the cycle time through the filter set is only 8 seconds. They
report values of optical depth $\tau$ for $\lambda=$ 340, 380, 440, 500, 675, 870, 940 and 1020 nm. 
They use the 940~nm channel for measuring water content, by comparing the 
solar flux seen there to that of adjacent bands. 

The AERONET Level 2 data have been selected to avoid days with obvious
clouds, and the data are processed in order to extract the various
components of attenuation, as described in \cite{Holben99}. 
In the descriptions that follow
we used the Level 2.0 AERONET data, obtained from {$\tt http://aeronet.gsfc.nasa.gov/new\_web/data.html$}. 

\begin{figure}
\plotone{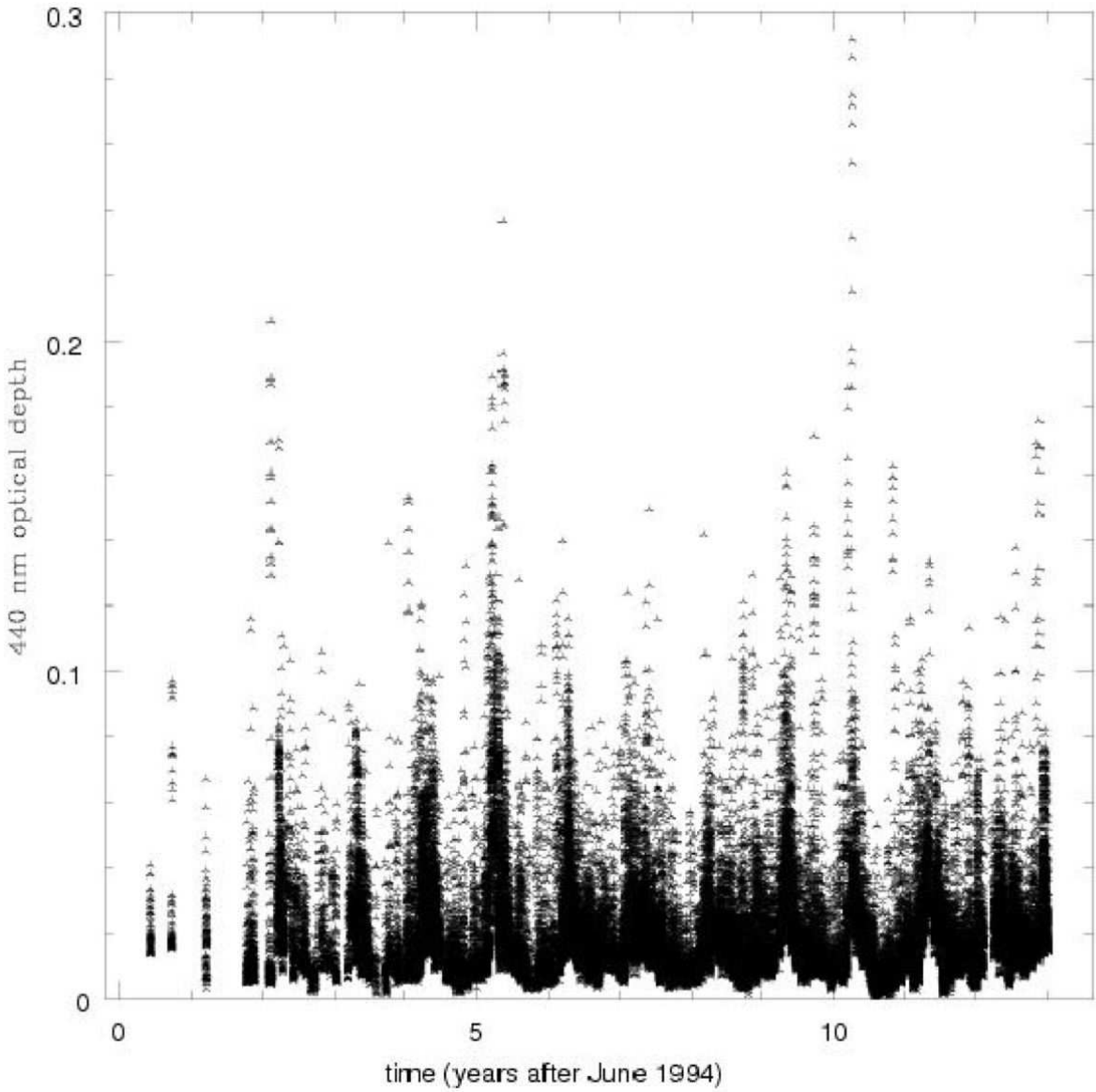}

\caption{Aerosol Optical Depth. The graph shows the daytime aerosol optical depth at 440 nm over time  
as reported from solar flux measurements by the AERONET system on Mauna Loa, near the site of the 
PanSTARRS-1 system. 
The vertical axis is optical depth $\tau$ where a fraction $e^{-\tau}$ is 
transmitted through one airmass. Time is in years after June 1994). 
There is clear evidence for seasonal cycles, as well
as considerable variation on short timescales. } 
\label{fig:t440_full}

\end{figure}

\begin{figure}
\plotone{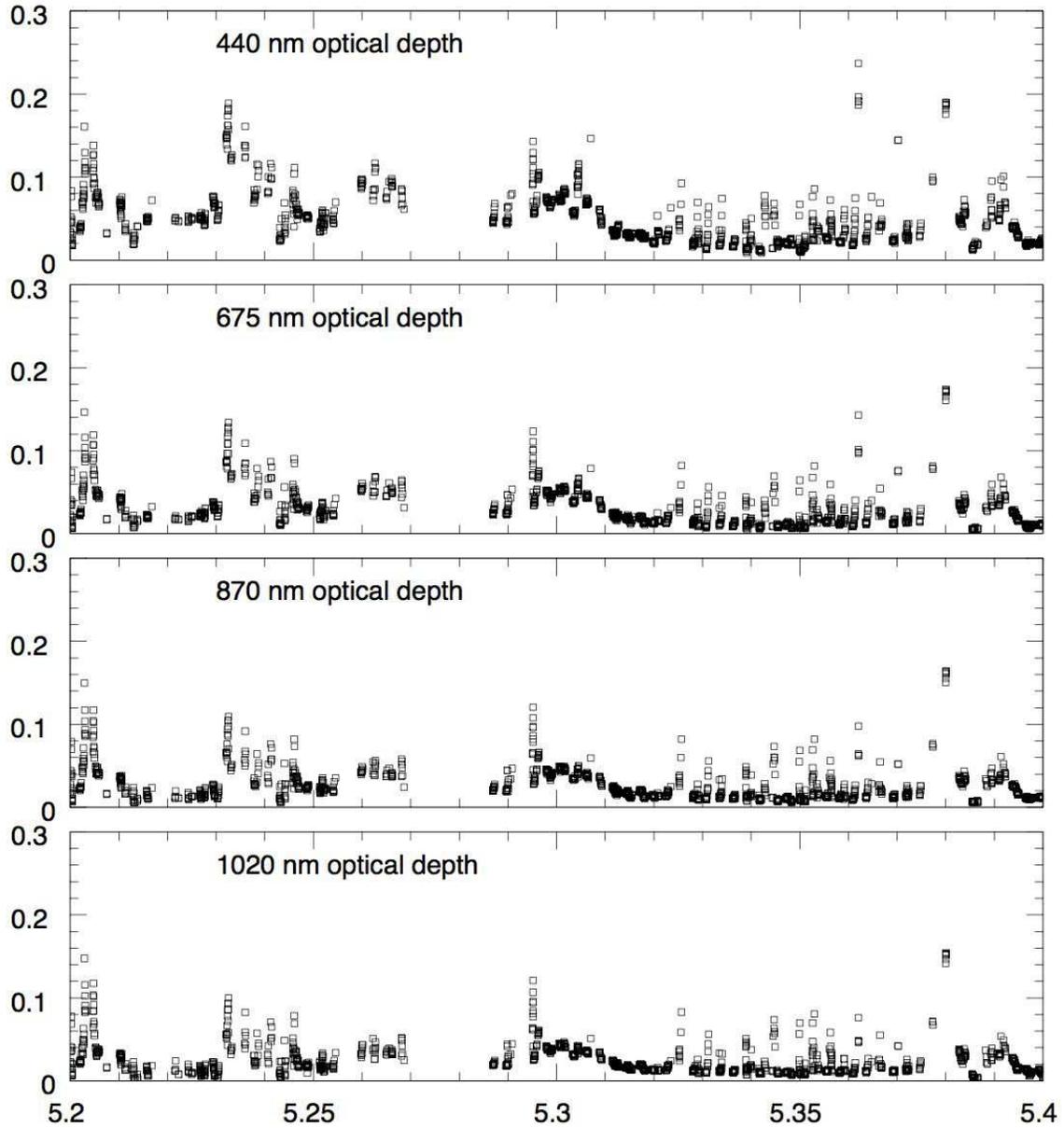}

\caption{Aerosol Optical Depths at Different Wavelengths. This plot shows in more detail the AERONET reported daytime Mauna Loa
aerosol optical depth at different wavelengths, vs. time. The plots span 
a period of 72 days. Note the variation in spectral dependence of the 
attenuation spikes.} 
\label{fig:depth_zoom}

\end{figure}

\begin{figure}
\plotone{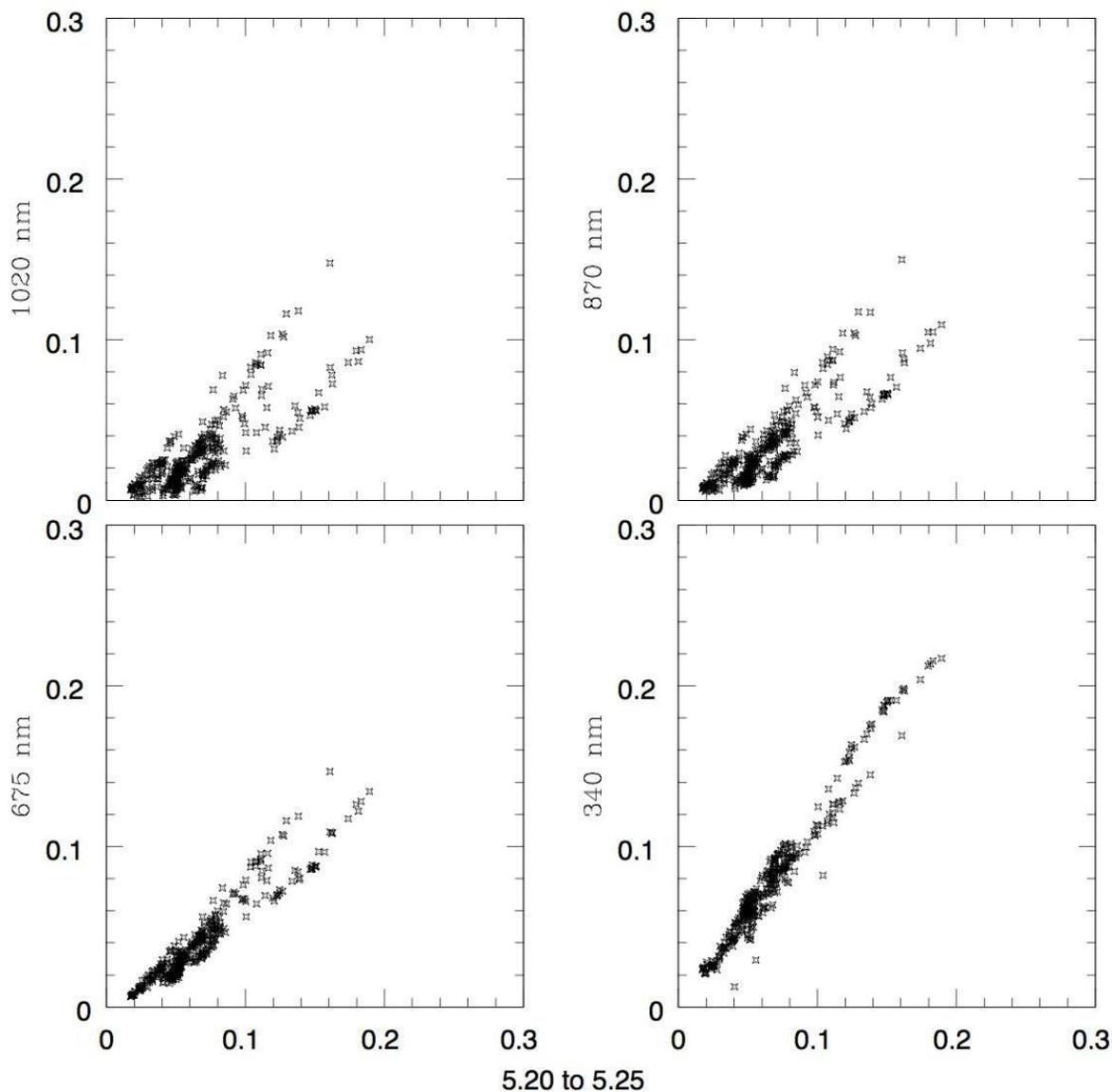}

\caption{An Extinction Excursion. 
These panels show the Mauna Loa AERONET attenuation at 1020 nm (upper left), 
870 nm (upper right), 675 nm (lower left), and 340 nm (lower right) 
plotted vs. the attenuation at 440 nm, for the period between 5.20 and 5.25
years of Figure 4. A wavelength independent change in transparency would 
generate a line with a slope of $m=+1$ in each of the panels. Aerosol scattering
would have slopes less than unity in all except the lower right panel. The 
data show evidence for both kinds of excursions in this time period.} 
\label{fig:quad1}

\end{figure}

\begin{figure}
\plotone{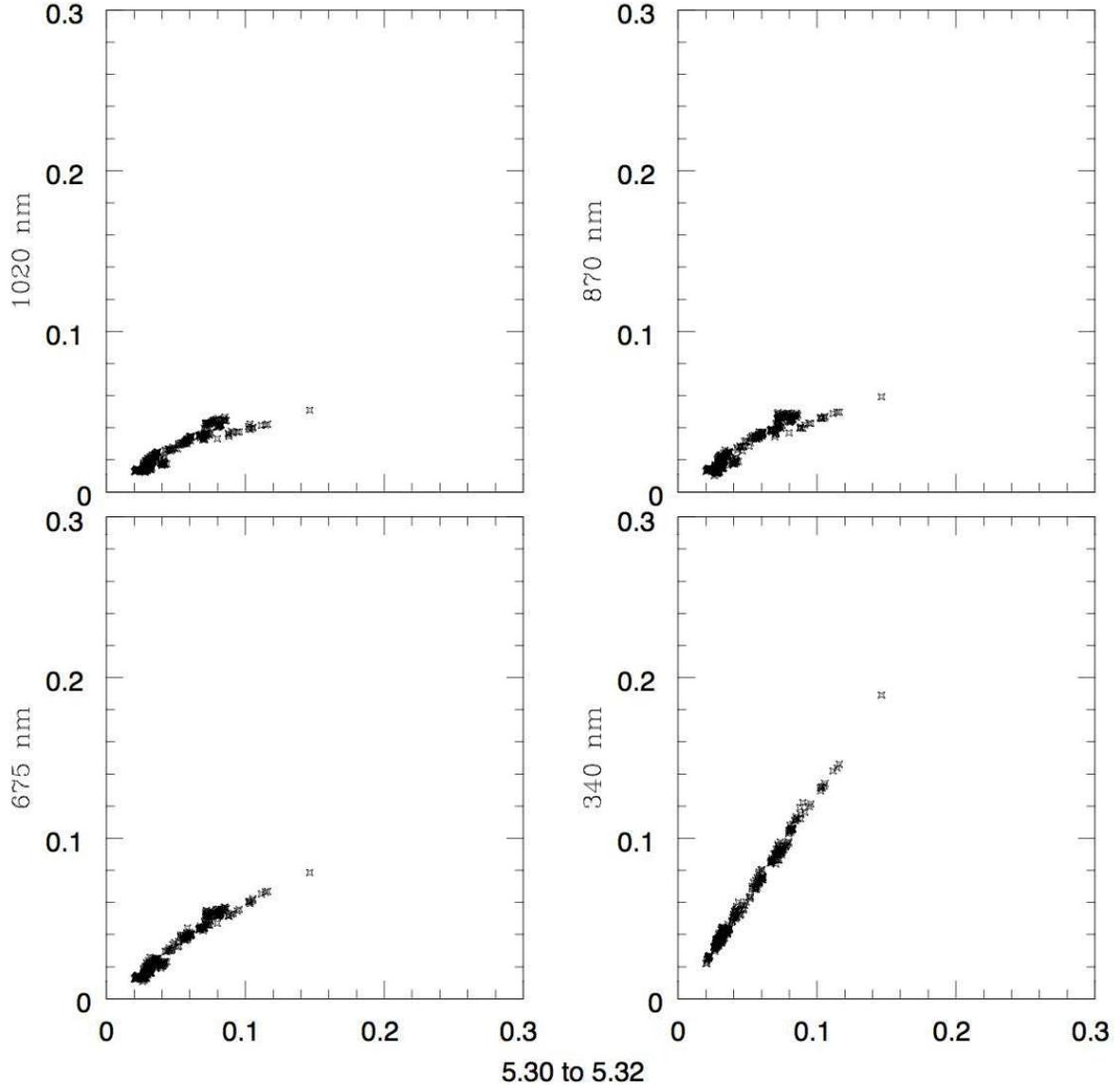}

\caption{An Extinction Excursion. Same panels as in Figure 4, but for the period between 
5.30 and 5.32 years. This attenuation spike appears to be predominantly
due to aerosols.} 
\label{fig:quad2}

\end{figure}

\begin{figure}
\plotone{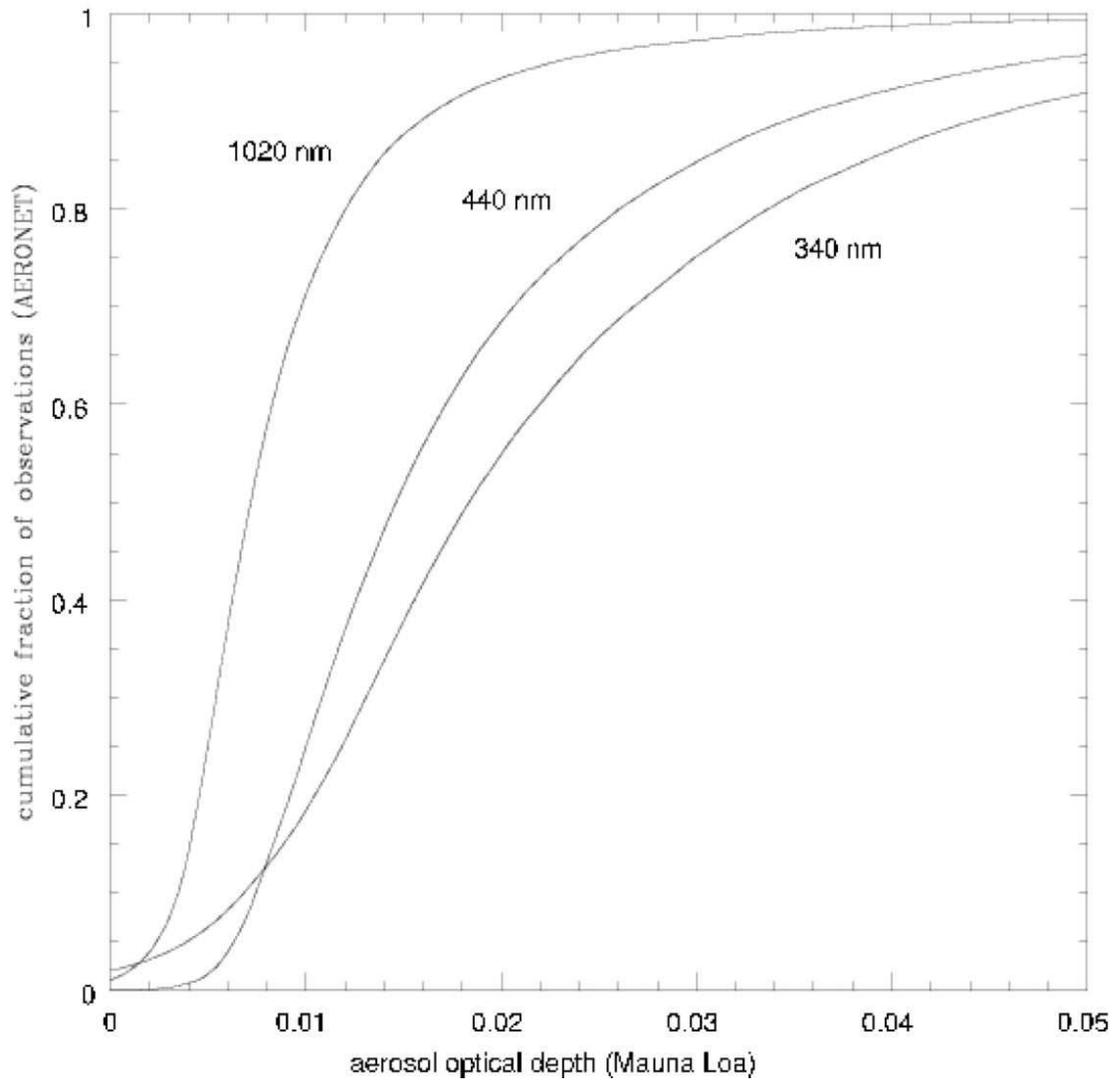}

\caption{Aerosol Optical Depth Statistics. Cumulative distributions of aerosol optical depth on Mauna Loa
are shown, 
from the AERONET observations over the period shown in Figure 2.} 
\label{fig:cumulative_aerosol}

\end{figure}

\begin{figure}
\plotone{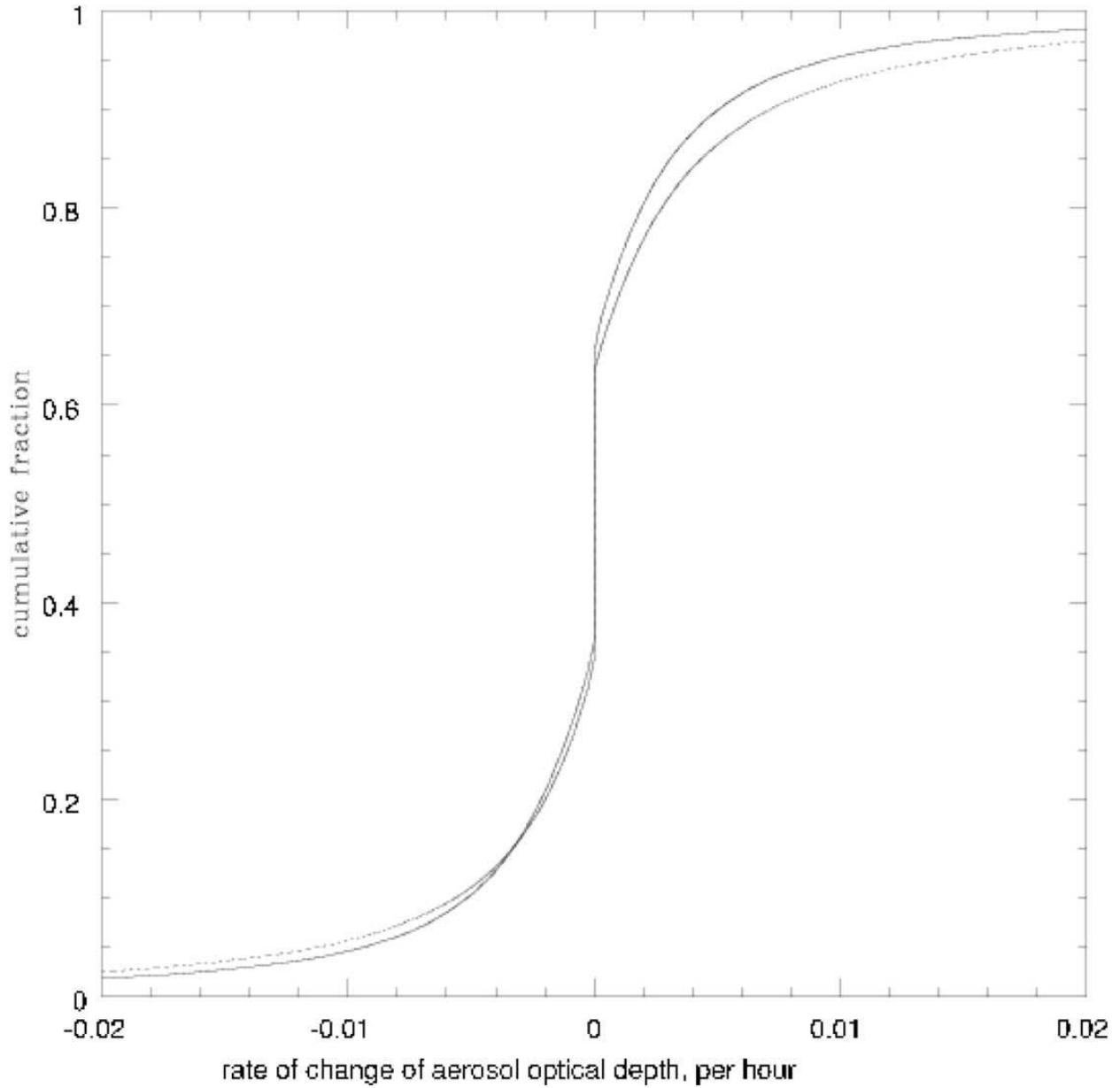}

\caption{Variation Statistics for Aerosol Optical Depth. Cumulative distributions of aerosol optical depth changes (per hour) 
for the AERONET data on Mauna Loa are shown. The solid line corresponds to 1020 nm
and the broken line to 440 nm.} 
\label{fig:cumulative_deriv}

\end{figure}

Figure~\ref{fig:t440_full} shows a 15 year record of aerosol optical depth at
Mauna Loa, at 440 nm. 
Figure~\ref{fig:depth_zoom} is an expanded view of a period spanning 0.2 years, and 
shows the aerosol optical depth fluctuations at 1020 nm, 870 nm, 675 nm, 
and 440 nm. Figures~\ref{fig:quad1} and~\ref{fig:quad2} show that different ``extinction spikes''
have different spectral character. Some extinction spikes are changes in 
transparency, with little wavelength dependence, while others clearly 
exhibit more attenuation at the shorter wavelengths. Figure~\ref{fig:cumulative_aerosol}
shows the cumulative aerosol optical depth distributions for 
the period of time shown in Figure~\ref{fig:t440_full}. The median aerosol optical 
depth values at 1020 nm, 440 nm and 340 nm are 0.007, 0.015, and 
0.018 respectively. Figure~\ref{fig:cumulative_deriv} shows the cumulative
distribution of variation in the AERONET aerosol optical depth measurements.

Measuring the extinction due to aerosols is one of our main challenges.  
It may be that daytime measurements, of the kind reported by \cite{Holben99}, 
could be used to estimate night-time aerosol scattering, but we need to explore 
the extent to which this is true. 
Most aerosols are low in the atmosphere. This, and the observed
short-term variation in aerosol optical depth, implies that we should expect
azimuth dependence of the aerosol transmission across the sky. 

\subsection{Molecular Absorption}

Absorption features in the spectra of atmospheric constituents produce a 
complex set of absorption bands and features. Ozone, water, O$_2$ and OH molecules
all contribute to this absorption. In fact, the atmosphere is essentially opaque in numerous
narrow regions, 
especially for $\lambda >$ 740 nm. 
A high spectral resolution determination of the molecular
atmospheric absorption above KPNO is presented in \cite{KPNO}. 

\subsubsection{Oxygen}

The strong absorption features at 690~nm and 760~nm due to O$_2$, the ``B'' and ``A'' bands in the nomenclature of Fraunhofer, 
are stable over time since they depend on the integrated Oxygen
content along the line of sight. The same arguments given above about the pressure dependence of
Rayleigh scattering apply here as well.  We therefore expect this component of atmospheric transmission, $T_{O_2}$,  to be axisymmetric about the zenith and stable over time, again with a slight pressure dependence. 

This picture is borne out by inspecting the corrections applied to spectrophometric data by \cite{Bessell99}. Figures 
1 and 2 of that paper clearly show much smaller residuals in the A and B bands as compared to water absorption regions.  

\subsubsection{Ozone}

\begin{figure}
\plotone{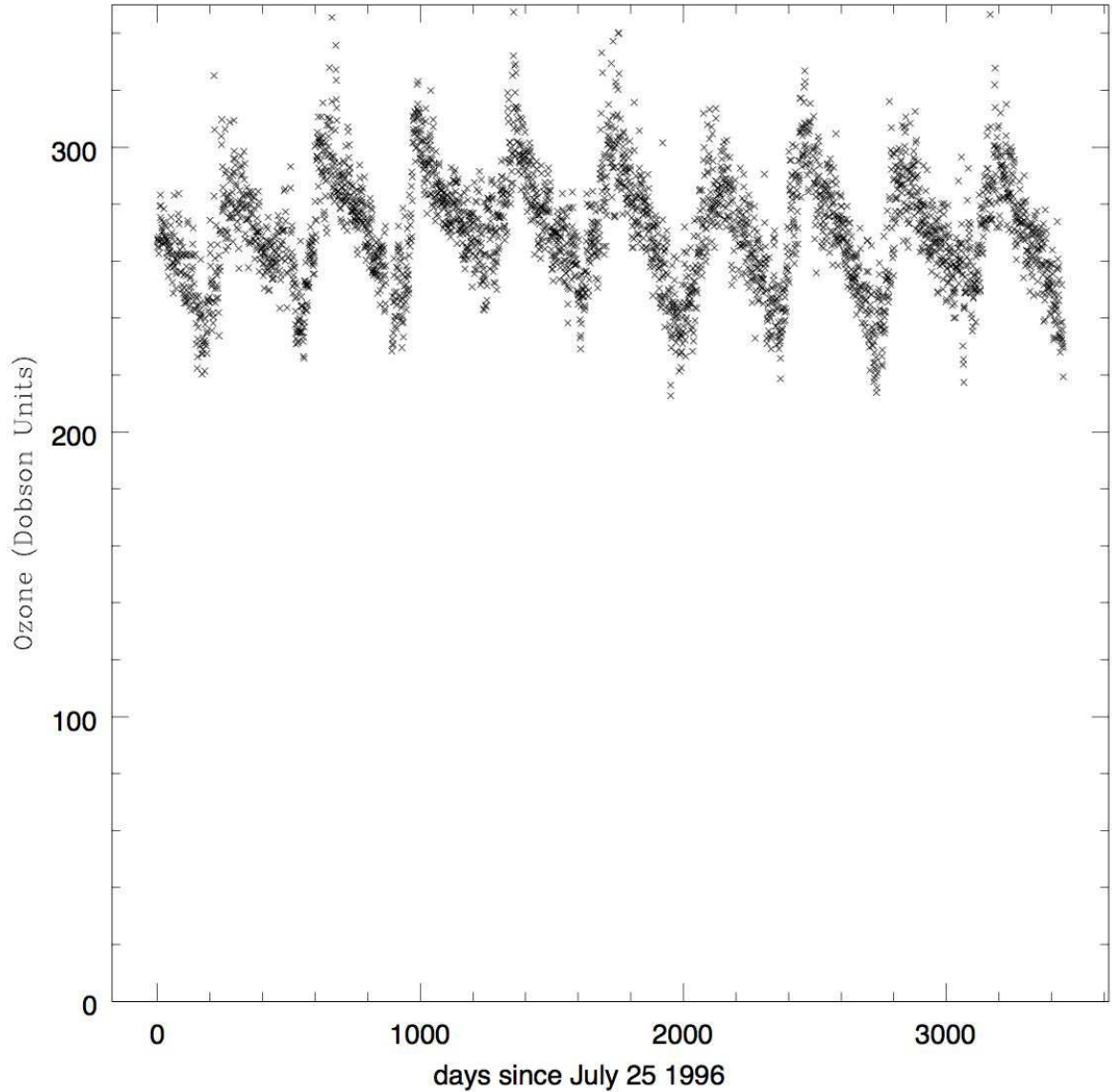}
\caption{Ozone Attenuation Variability. This plot shows the evolution of the ozone content of the 
atmosphere vs. time above Hilo, HI, as measured by the EarthProbe TOMS 
satellite-borne instrument. The
y axis is in units of Dobsons. Each Dobson unit is equivalent to a thickness of  
0.01 mm of ozone at STP. There is clear evidence of annual cyclic variation
at the $\pm$25\% level about the mean value. Remote sensing data such as these
can be used to determine the optical attenuation due to ozone without needing 
any ground-based measurements.} 
\label{fig:ozone}

\end{figure}

The opacity of ozone is responsible for the total loss of atmospheric transmission below 300nm. The Chappuis band of ozone influences transmission for 500~nm~$< \lambda <$~700~nm, 
with an attenuation of a few percent at 600~nm. 
The measurement of ozone in the atmosphere is of great interest due to 
its important role in the Earth's radiation balance and climate change. 
This has led to the development and deployment of 
sophisticated space based instruments that are optimized for the 
determination of the ozone content of the atmosphere. These data
sets can be used to determine the ozone above observatories. 
Satellite measurements of the ozone content versus time over Hilo, HI
are shown in Figure~\ref{fig:ozone}. These illustrative data were obtained from the
TOMS instrument \citep{Jaross03} 
on board the EarthProbe satellite, and we obtained the 
data for Figure~\ref{fig:ozone} from {\tt http://toms.gsfc.nasa.gov/eptoms/ep\_ovplist\_a.html}.

The determination of ozone content in the atmosphere above the 
observatories can be obtained from satellite remote sensing data
with sufficient accuracy and temporal resolution for our purposes. 
Since the attenuation due to ozone is only a few percent, we need
only know the ozone content with a fractional precision of around ten percent.
\cite{Eck99} state that a 50\% change in ozone content would perturb
the optical transmission by 0.0036, 0.0045 and 0.0063 at 340 nm, 
500 nm and 657 nm respectively. 
 
We expect therefore to be able to exploit data obtained from satellites to constrain nightly variations in attenuation due to ozone, 
verified by direct measurements of atmospheric throughput on and off the well-known spectral regions of ozone attenuation.


\subsubsection{Water}

\cite{Frogel} presents evidence for variation of atmospheric attenuation due to water, measured in the IR, and shows month-to-month and longer term variability. \cite{Roosen77}  show that surface humidity
measurements only poorly correlate with the upper atmospheric water burden. Attempts \citep{Bessell99}
to make a statistical correction to spectra for the absorption due to water
have significant residuals, compared to oxygen A and B bands, which is evidence for 
significant variability. The PanSTARRS
filter set includes a $y$ band that takes advantage of the enhanced red 
response of high resistivity CCDs, and the break between the 
PanSTARRS $z$ and $y$ bands has been selected
to avoid the large water absorption feature at 950 nm. Nevertheless there are 
H$_2$O absorption features, as shown in Figure~\ref{fig:photon_transmission}, that impact other spectral regions. 

\begin{figure}
\plotone{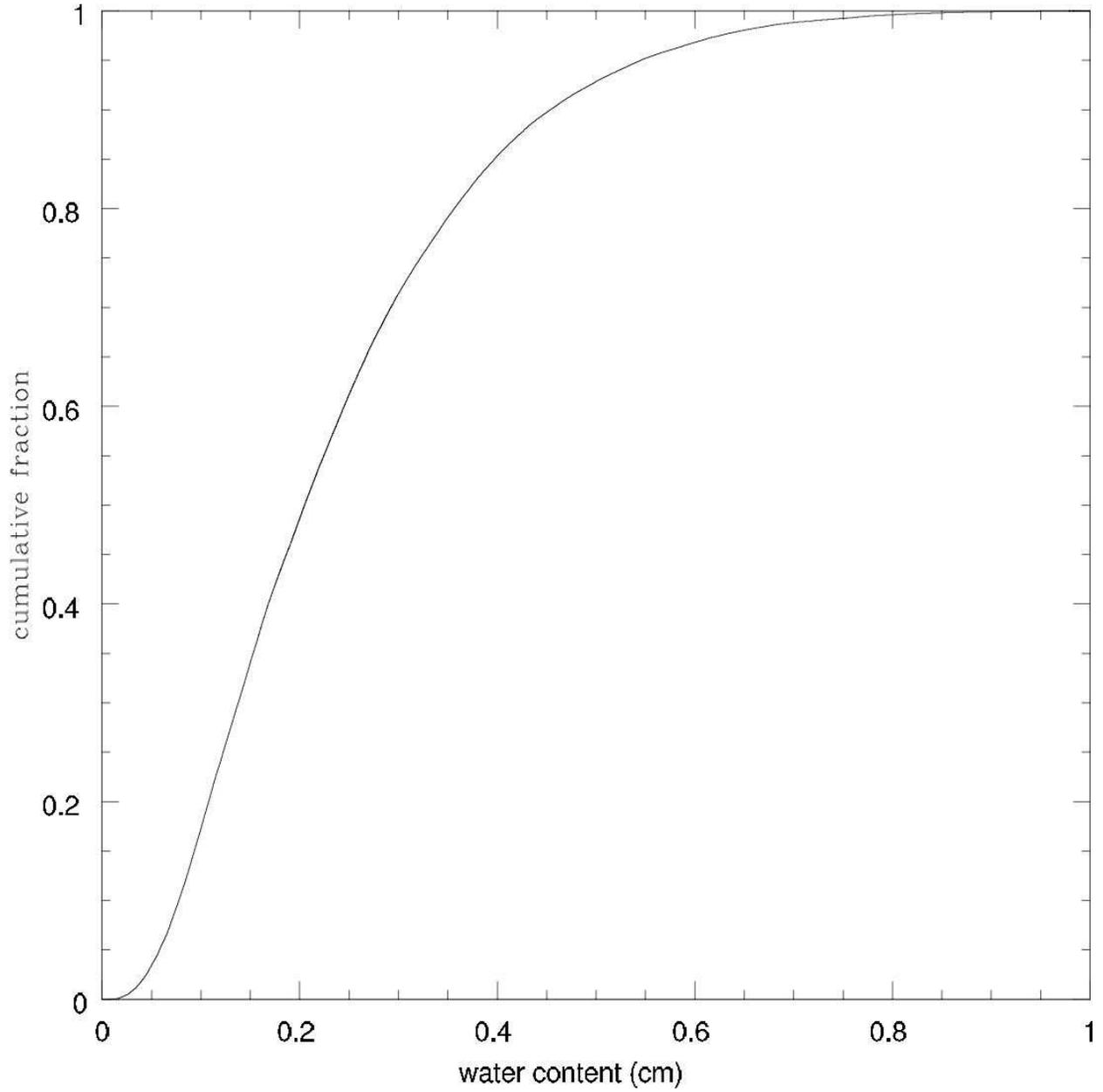}

\caption{Cumulative Distribution of precipitable water content (cm) above Mauna Loa, from the 12 year AERONET data set. } 
\label{fig:cumulative_water}

\end{figure}

\begin{figure}
\plotone{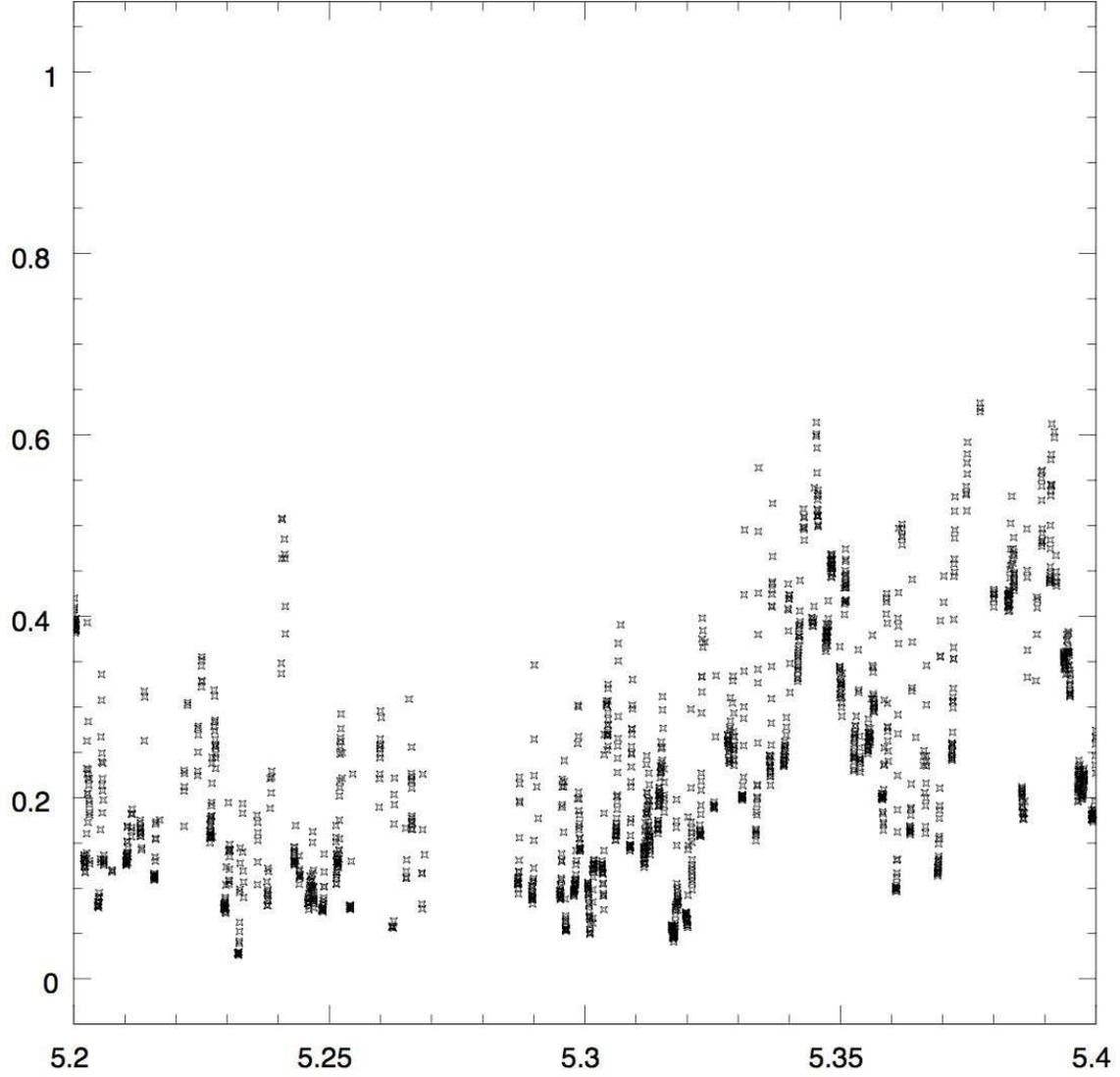}

\caption{Time evolution of water content of the atmosphere at Mauna Loa.
This plot of water content (in cm) vs. time covers the same interval as
the plots in Figure~{\protect \ref{fig:depth_zoom}}.} 
\label{fig:watervt}

\end{figure}

Figure~\ref{fig:cumulative_water} is a cumulative distribution of the measured water content
in the atmosphere over a ten year period, from the AERONET 
station on Mauna Loa. The units used are cm of precipitable water. 
Figure~\ref{fig:watervt} shows the temporal evolution of the water content over the 
same period shown for the aerosols in Figure~\ref{fig:t440_full}. 

\cite{Campanas07} present an interesting comparison of measurements of atmospheric water
content using both 255 GHz radiometery and high resolution optical spectroscopy. Their paper
contains a valuable review of the physics of optical absorption by water vapor, and 
demonstrates good agreement between the results obtained with these two techniques. 
They also show data with clear indications of short term variability in 
atmospheric water, over the course a night. The use of RF radiometry for the 
determination of water absorption in the optical clearly merits further consideration.

The evidence indicates that we should expect substantial temporal fluctuations in the optical 
absorption due to water, and this implies a potential azimuth dependence as well. 
We therefore conclude that optical attenuation due to water in the atmosphere is our second 
main challenge, with a complex behavior of both $\alpha_{water}(\lambda)$ and $T_{water}(\lambda, \hat{z}, t)$. 

\subsubsection{Clouds}

Water droplets and ice crystals also attenuate the transmission of
optical radiation through the atmosphere. The standard assumption is that these
objects that make up clouds are sufficiently larger than the wavelengths in question that
the scattering is wavelength independent, so that clouds are ``grey'' scatterers. 
\cite{EH01} achieved Poisson limited performance in differential photometry through high cirrus, lending
credence to the idea that clouds are grey absorbers. 
Clouds can be detected through their emission (e.~g.~with a camera operating in the 10 micron band), 
by optical attenuation (through their effect on survey photometry), or in reflection
(using LIDAR, see \cite{Auger1}). 

A thorough examination by \cite{Ivezic07} of drift scanned
SDSS data from repeated imaging of stripe 82 has also shown little dependence of 
photometric residuals with color, even through many magnitudes of extinction from clouds, and even when comparing $u$ and $z$ band fluxes. 
These authors place an upper limit of 0.02 mag of zeropoint uncertainty for photometry obtained
through one magnitude of extinction from clouds. It is important to distinguish
``astronomical point source'' transmission through clouds, which we take to be the 
unscattered component, from the net radiation transfer of flux from the sun, 
as described in \cite{Kylling97}.

One might imagine that a survey would need at least one visit to each field under 
highly photometric conditions in order to achieve a uniform all-sky zeropoint in each 
passband. The SDSS experience suggests that the location of the stellar locus, in color-color
space, might be used to make zeropoint corrections for those bands that were imaged
through clouds. This of course assumes that the stellar locus has no dependence on Galactic
coordinates. 

PanSTARRS is planning to use a modest aperture photometric monitor that places
the entire survey field of view onto a single CCD array. This imager will be used to monitor the 
flux from Tycho catalog stars in the field \citep{MagnierPS}. Using these objects as flux references will 
allow for the detection of even small amounts of grey extinction, and scatter in the 
flux differences across the field will be indicative of variable cloud across the field of view. 

Contrails from aircraft are a nuisance that the astronomical community has not really faced
up to. These could in principle be detectable by searching for residuals in the stellar locus
that are linear across a frame, and excising or correcting the afflicted data.  While we are not aware
of any project that has done this, it should be straightforward. An approach akin to the
excision of satellite tracks could be implemented, for example, based upon
photometric residuals that lie along a common line in an image.

\section{THE ANGULAR AND TEMPORAL CORRELATION FUNCTIONS OF ATMOSPHERIC TRANSMISSION}

The different contributions to atmospheric optical attenuation have 
different expected azimuth, zenith angle and time dependencies.  
Short term (during the night)
time variability must necessarily be accompanied by variation in the azimuth and zenith
angle dependence. The converse is not necessarily true, as we can imagine
non-axisymmetric configurations of the atmosphere that are stable over a night. 

\cite{Anderson03} have published the two-point correlation functions in space and time
for aerosol scattering, but their data were not necessarily representative of an astronomical site. 
Nevertheless, they show evidence for the aerosol scattering component being highly 
correlated (correlation coefficient $r>0.8$) over time scales of many hours, and length scales of 
tens of km. 

We took the time series AERONET data and determined the rate of change
of aerosol optical depth at 1020 nm and 440 nm. Figure~\ref{fig:cumulative_deriv} shows the cumulative distribution of the rate of change of the measured attenuation, 
in units of (optical depth change) per hour. 
It would appear that hourly measurements would have optical depth 
changes of under 0.01/hr for more than 90\% of the time, assuming
nighttime variation is similar to the daytime AERONET data set. Taking 
10 m/s as a typical wind
speed at the typical aerosol height of z=4 km, in one hour the atmosphere will have translated by
36 km.  This implies that the aerosol burden is correlated over at least
this angular scale. We can therefore be optimistic that the aerosol attenuation
will be correlated across the sky, for an hour at a time. 

\begin{figure}
\plotone{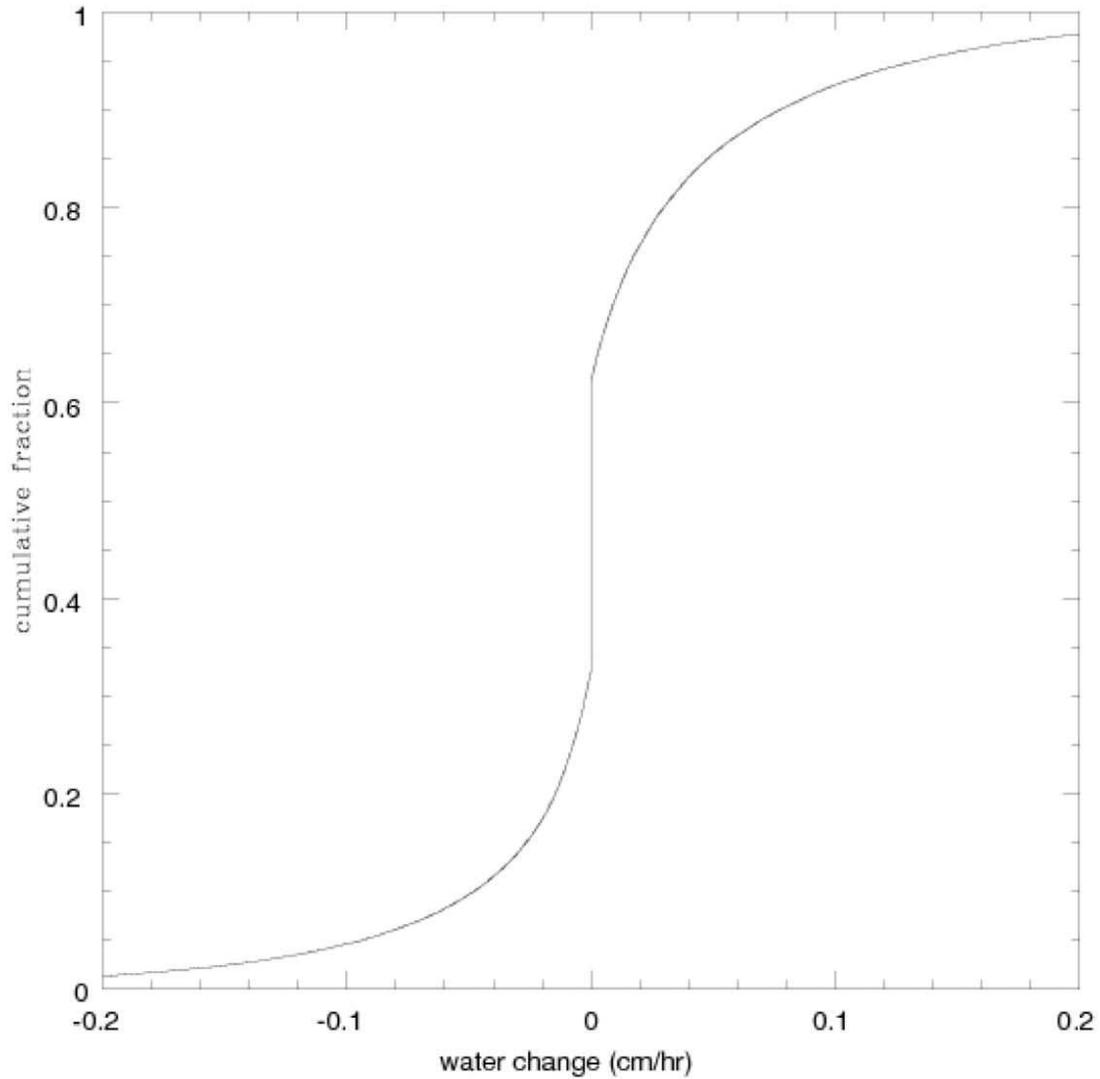}

\caption{Cumulative plot of rate of change of AERONET's derived 
precipitable water content (cm per hour). For a typical value of 0.2 cm
this implies that measurements more frequently than hourly are 
required to track 10\% changes with confidence. } 
\label{fig:cum_water_deriv}

\end{figure}

A similar analysis for the rate of change of water is shown in 
Figure~\ref{fig:cum_water_deriv}. This appears more demanding. 
The water column can change by 10\% in less than an hour, implying
an azimuth variation as well. Figure~\ref{fig:watervt} clearly shows
as much as a factor of 4 variation in water content of the atmosphere at Mauna Loa. 
Since we would expect the optical depth due to water to scale as the square root
of the water content in the air, this implies a factor of two fluctuation in optical
attenuation due to water. Referring back to Table 1 we note that water
absorption accounts for 3\% and 5\% attenuation in the $z$ and $y$ bands,  
respectively, at the zenith. This in turn implies that we should expect 1-2\% variations in 
zeropoint in these bands due to atmospheric  water content variation. This effect will 
of course increase with zenith angle.  
  
An important contribution by \cite{Ivezic07} is their measurement of the structure function of clouds. These authors state that the zeropoint offset between two regions
is proportional to both the angular separation $\theta$ and the cloud's 
extinction $A$, with $\Delta m \sim (0.02$ to $0.10) \times A \times \left(\theta/1^o\right)$. 
This implies that local relative photometry can be extracted through clouds
over surprisingly large areas, in agreement with the results 
from \cite{EH01}.  

\subsection{The implications of a wide field of view, and aperture considerations}

Much of the methodology currently used for correcting for atmospheric
extinction is carried forward from the era of single-pixel detectors, but
a distinguishing feature of the next generation of sky surveys is their 
wide field of view and Gigapixel CCD arrays. 
The PanSTARRS and LSST imagers will span 3.0 and 3.5
degrees, from corner to corner. Even at a modest zenith angle of 60 degrees, 
the span in airmass across the LSST imager will range from 1.9 to 2.1! 
Taking the extinction values from Table 2, the {\em differential} extinction across the field will be 0.10, 0.04, 0.02,  and 0.01 magnitudes
in $u,g,r,i $ and $z$, from this effect alone, at $\chi=$2 airmasses.

Our measurements are under the influence of attenuation processes 
that exhibit different airmass dependences, so we cannot simply assign 
a single effective atmospheric transmission curve to each survey image, 
but instead we should calculate what amounts to a wavelength dependent
illumination correction across the field of view. 

The atmospheric column traversed by the light from different objects depends
upon their separation and on the aperture of the telescope. 
Using the same arithmetic that applies to multi-conjugate
adaptive optics, the beams from two sources separated by an angle $\theta$ traverse distinct regions of the upper atmosphere only above a height $h>D/sin(\theta)$, 
where $D$ is the telescope aperture. This means that for LSST, with $D=8.5$~m, 
the atmospheric transmission below $h=15$ km is ``common mode'' for objects 
closer than 1 arcminute on the sky.  Aerosol attenuation arising from 
a static layer at z$\sim$4 km would be common over a focal plane region spanning 3 arcmin on LSST, and  0.6 arcmin for PanSTARRS.
Local atmospheric perturbations in the first few
hundred meters above the observatory are common mode across the entire LSST
field. The dynamics of the atmosphere greatly suppress possible variation 
across the field of view, however. A 10 m/s wind will drag hundreds of meters
of atmosphere across the telescope aperture over a typical survey exposure time of 
15 or 30 seconds. This serves to homogenize the time-averaged atmospheric
profile through which the system is imaging, at least in one dimension, across the entire field of view.

\subsection{Measurement Priorities}

Based on the above considerations, we can expect the Rayleigh and 
O$_2$ components of atmospheric transmission to be well behaved, 
with minimal variation from night to night and little azimuth dependence. 
The spectral shape of these mechanisms is also essentially 
time independent. The aerosol term, on the other hand, clearly 
shows evidence for temporal variation and spectral evolution, and 
is difficult to measure. Water absorption has a fixed spectral profile but
its strength is highly time variable. Clouds are grey absorbers at the
1\% level, with significant temporal variation. Ozone absorption is time
variable but can be measured from satellites. 

Our priorities are therefore the determination of aerosol scattering and
absorption due to water molecules. 
  
\section{SPECTROSCOPIC MEASUREMENTS OF ATMOSPHERIC TRANSMISSION}

There is a long tradition in astronomy of using spectrophotometry to
determine the spectral energy distribution of celestial sources of
interest. An integral part of the analysis of these data is to
compensate for atmospheric attenuation. This naturally provides an
opportunity to determine the wavelength dependence of atmospheric
transmission, within the spectral resolution limitations of the
apparatus. \cite{Bessell99} devised a technique to disentangle the
spectral structure of the atmosphere from that of the celestial
source. \cite{Stritzinger05} used a combination of their
spectroscopic observations and stellar models to produce a tabulation
of extinction values versus wavelength, derived from their measurements
of spectrophotometric standard stars. We caution, however, that their
tabulated extinction values do not include molecular absorption
effects.  We can consider the smooth extinction data of
\cite[][; their Fig.~2]{Stritzinger05} to be a good complement to the
high spectral resolution absorption atlas of \cite{KPNO}.
 
In principle, high precision spectrophotometry seems like the ideal
way to measure the atmospheric attenuation profile. Measuring the
Rayleigh, aerosol and ozone attenuation does not require much spectral
resolution (on the order of $\sim1$\,nm), but does benefit from a large
spectral range (300\,nm to 1\,$\mu$m) and requires high precision
spectrophotometry. The strength of the molecular absorption features
can be expressed in equivalent widths (as commonly done amongst
stellar spectroscopists), and since this is a spectrally local
differential measurement, broadband flux precision and knowledge
of the source spectrum are less of a
concern.

One major advantage of the spectroscopic approach is that the
measurements at all wavelengths are made simultaneously. This helps
distinguish spatial from temporal variation in transmission.  

There are two possible approaches to using spectroscopy for the
determination of atmospheric extinction. If the spectrum of the source
and the wavelength dependence of the instrumental sensitivity are well
known, a single measurement of the spectrum of the source suffices to
determine the atmospheric attenuation vs. wavelength. On the other
hand, measurements obtained at a diversity of airmasses can be used to 
distinguish between the airmass-independent aspects (source spectrum and
instrumental throughput) and the airmass-dependent atmospheric
attenuation, liberating us from having to know the source
spectrum. This second approach of course is susceptible to systematic
errors due to any time dependence of the atmospheric extinction. 

Drawbacks to the spectroscopic approach include the loss of signal to
noise, compared to the imaging techniques described below, due to
dispersion. There are also non-trivial instrumental
challenges. Grating dispersion elements have a large variation in
throughput with wavelength, and maintaining a high signal to noise ratio across a wide
spectral range is difficult due to limited CCD response at the limits
of the targeted wavelength range. Prisms have higher dispersion in the
blue than the red, which is the inverse of what we'd like to have, 
although this is in part compensated by the loose constraint on
spectral resolution.

A slitless spectrograph is essential for high precision measurements
in the blue at high airmass, but guiding errors can produce a
systematic offset between the wavelength solution from arc lamps and
the actual spectrum of the object. Nevertheless, the object
spectrum itself can be used to determine a wavelength solution
through identification of lines commonly found in stellar spectra
(such as the Balmer series of hydrogen).

Extracting information about the distinct attenuation processes
described above invariably involves comparing the fluxes in certain
spectral regions, and (for the absorption lines) determining
equivalent widths. This means the spectrometer is essentially being
used as a simultaneous narrowband imager. 

We have begun observing campaigns at Haleakala and CTIO 
designed to address these issues.
The first steps have focused on assessing our ability to detect the signatures of the elements
of atmospheric extinction in modest-resolution (R $\sim$ 400) stellar spectra.
Multiple stars are observed as they progress across the night sky.
Spectral features whose strengths vary with airmass are extracted by fitting to templates
of atmospheric extinction computed with models.
By use of appropriate patterns of stars on the sky one can attempt to disentangle the 
spatial and temporal characteristics of the major contributors to extinction.
Results from these studies will be subjects of future papers.

We carried out spectroscopy of a variety of standards stars of spectral class O-F
at the Hawaii 2.2~m on Mauna Kea, using the SNIFS spectrograph \citep{SNIFS04}, and on the CTIO 1.5~m with their
Ritchey-Chr{\'e}tien spectrograph. We will report on our results
in subsequent publications, but as described in section 7.1 our preliminary reductions illustrate the
diagnostic power of spectroscopy in conjunction with transmission models for determining the airmass-dependence
of atmospheric attenuation. 

\section{PHOTOMETRIC MEASUREMENTS OF ATMOSPHERIC TRANSMISSION}

To measure extinction using imaging systems, astronomers have traditionally used the same broadband imaging system used for their program targets to
observe standard stars.  This automatically avoids the ``color terms" required when transforming between photometry from different telescopes, cameras, or filters. If a set of comparison stars with spectra identical to the program objects can be found in the same field, this can potentially deliver high photometric precision. For PanSTARRS and LSST this approach is impractical, due to the time
needed to change filters. 

A number of programs have been undertaken to characterize the behavior
of atmospheric transmission. At the risk of oversimplifying, the astronomical community has turned their instrumentation to this task, typically using standard
astronomical passbands. The atmospheric sciences community, on the
other hand, has been 
acquiring data with filter sets that are optimized for understanding the 
properties of the atmosphere. Most of their imaging is done in the daytime, 
however.

Schuster et al (2002) describe a campaign of attenuation measurements using 13 narrowband filters than span 330 nm $< \lambda <$ 630 nm.  However, these filters were chosen for measuring the properties of stars, not the Earth's atmosphere.

\subsection {A Simultaneous Multiband Stellar Photometer to Measure Attenuation}

We have built and are now testing a dedicated, simultaneous multi-narrowband imaging photometer (High et al. in prep.).  The system uses a mask at the optical aperture, onto which wedge prisms and narrowband filters are mounted, in front of a fast camera lens.  We put these on a commercial deep depletion CCD camera (Pixis 1024BR, from Princeton Instruments) to achieve the same quantum efficiency we expect from PanSTARRS and LSST, and take pictures of bright stars.  A prism wedge offsets the angle of light rays from a given star just in front of the aperture, resulting in an offset but still localized stellar image at the focal plane.  By orienting each wedge differently and filtering their light independently, we produce an array of PSFs from a given star at different wavelengths in a single exposure.

The filter set is akin to that used for daytime solar photometers that are used
to characterize the atmosphere. We chose the central wavelength of one filter to coincide with the main water absorption feature at 950 nm, and another just off the feature at 1000 nm.  Another filter coincides with the narrow oxygen absorption band at 760 nm, and the other 3 are positioned where the Rayleigh/aerosol scattering dominates.  All our filters have FWHM of 10 nm.  This filter set allows us to probe the main, narrow absorption bands, and the Rayleigh/aerosol scattering components. Using a single
common shutter makes this a clean differential measurement. 

\begin{figure}
\plotone{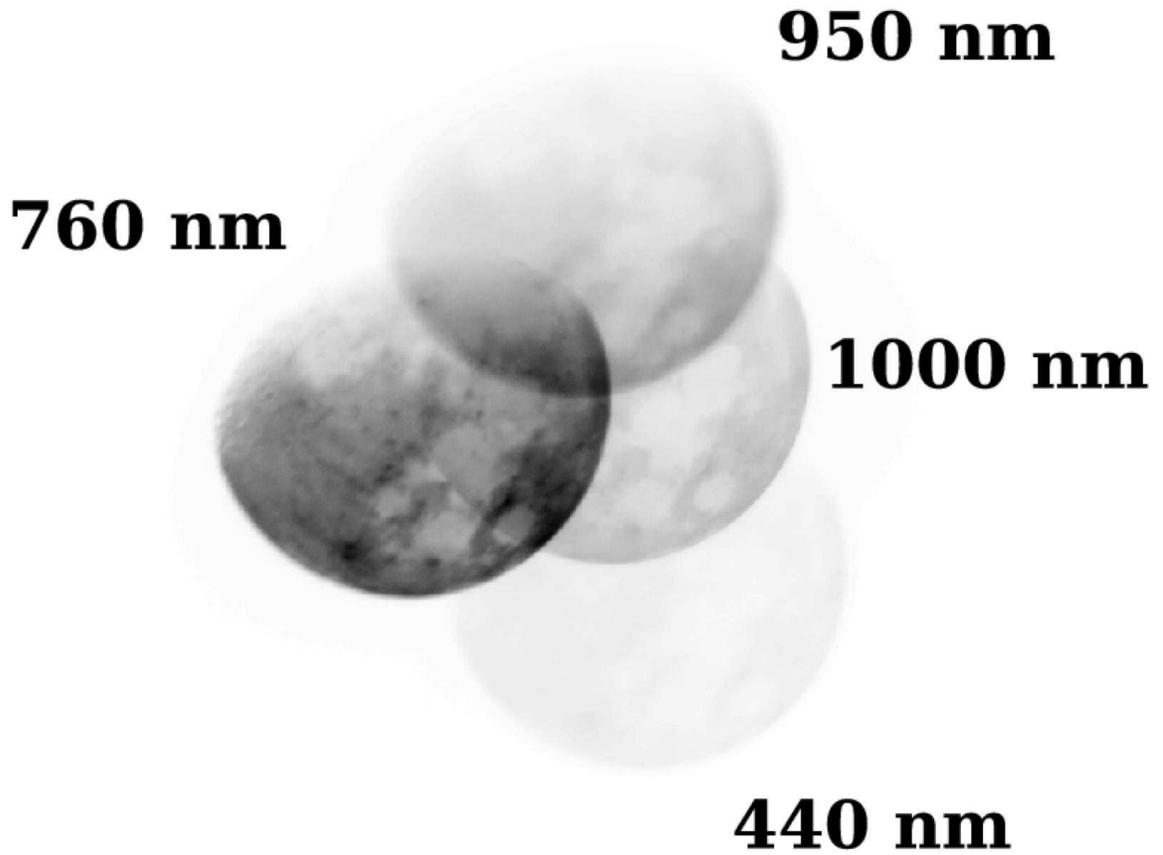}

\caption{Image of the moon obtained simultaneously in four passbands, 
as part of our development of an simultaneous multiband imaging instrument designed to measure
atmospheric transmission in real time. Images of a point source would be
well separated and the flux differences can be used to deduce attenuation.} 
\label{fig:moon}

\end{figure}

\begin{figure}

\plotone{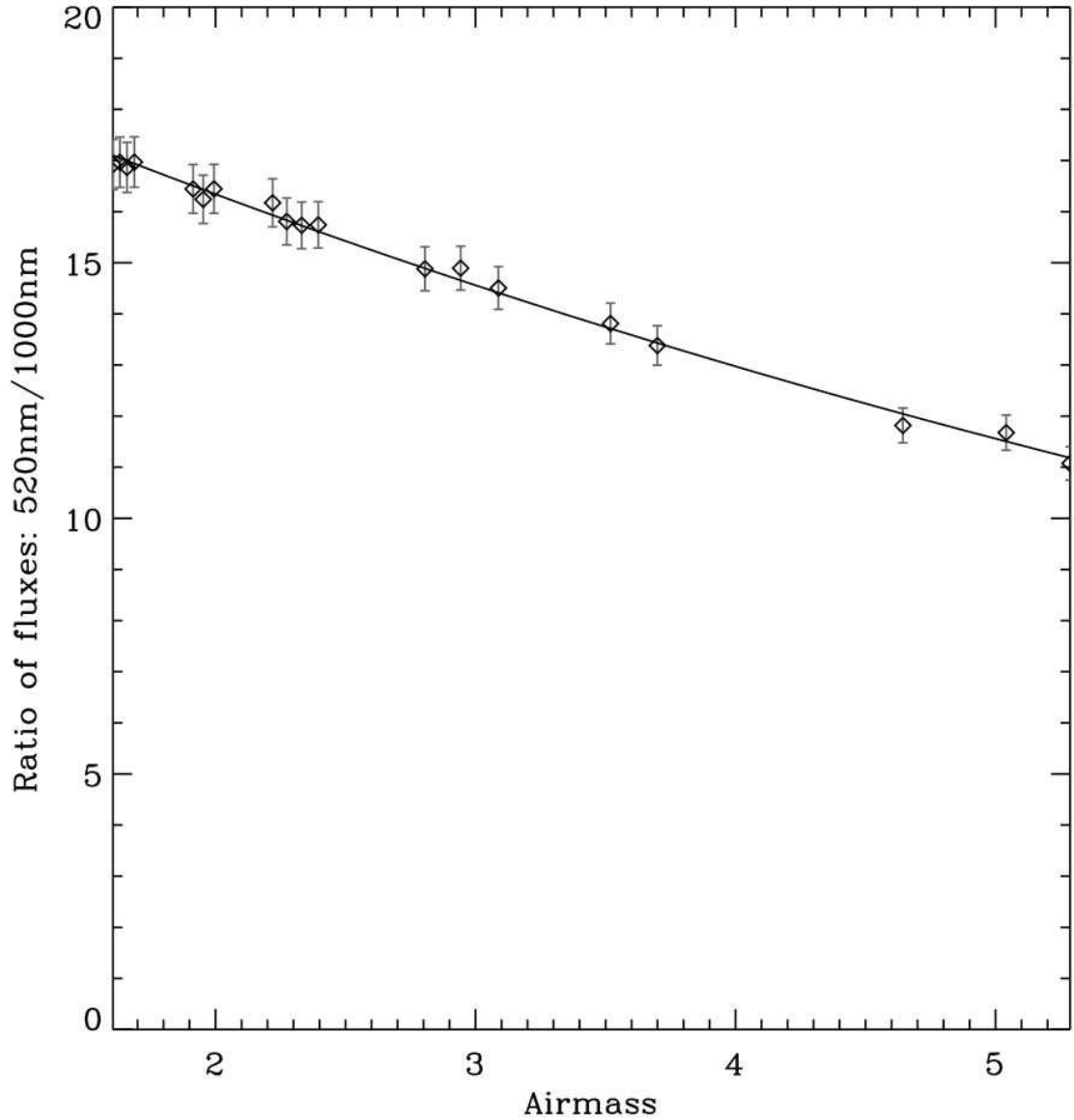}

\caption{Aerosol Attenuation Data from Simultaneous Multiband Imaging. This plot shows preliminary relative attenuation results we obtained 
using the multiband imager, from CTIO. The vertical axis is the ratio of transmission at
440 nm to that at 1000 nm, plotted vs. airmass. The two fluxes were obtained simultaneously
using the instrument described in the text. We consider this a promising technique to 
determine attenuation from both aerosol and molecular processes.} 
\label{fig:rainbow}

\end{figure}

A test image from our simultaneous multiband imager is shown in Figure~\ref{fig:moon}, 
 and a representative profile of optical transmission versus airmass is presented in 
 Figure~\ref{fig:rainbow}. 

Such an apparatus is compact and largely made from commercially available parts.  During operation, it leaves the scientific program uninterrupted.  The data would later enter directly and quantitatively into the atmospheric attenuation model during analysis.  It is therefore an interesting potential alternative to the traditional sequential broadband imaging methods.

\section{ACTIVE INTERROGATION OF THE ATMOSPHERE: LIDAR}

The atmospheric science community has long used LIDAR (Light Detection and Ranging) to probe the properties of the atmosphere. The book by \cite{Measures84} is a comprehensive reference for this technique. A LIDAR system sends pulses of light up through the atmosphere, and measures the intensity of backscattered light as a function of time. The difference in light travel time to different altitudes provides a method for probing the vertical structure of the atmosphere.

Light pulses emitted into the atmosphere can be scattered elastically or inelastically. The latter process results in a wavelength shift in the scattered light after internal degrees of freedom are excited in the scattering particles. Inelastic scattering has a smaller cross section, so the detection of these events requires some combination of a more intense light source and a detector with a larger aperture than is necessary to measure elastic scattering. The benefit of measuring the inelastic channel is the ability to isolate the type of scattering particle by measuring only the light shifted by specific wavelengths corresponding to quantum mechanical transitions for a chosen molecule such as N$_2$ or O$_2$, for which the vertical density profiles are well known. These constraints on the probability of scattering from the inelastic channel allow for direct measurement of round-trip extinction. 

LIDAR systems exploiting both the elastic and inelastic channels have been used for decades to study atmospheric properties. \cite{Melfi72} gives an 
overview of inelastic (Raman) techniques, and  \cite{Vaughan93} describe using inelastic scattering to obtain precise temperature 
profiles in the atmosphere.   \cite{ALE06} and \cite{Dawsey06} have described their program of developing an eye-safe elastic-channel single wavelength LIDAR system for the determination of atmospheric extinction of astronomical sources, and we look forward to results from this system once it is deployed. The 
use of multiple elastic-scattering LIDAR systems to carry out what amounts to a tomographic measurements of aerosols above an observatory is described
in \cite{Auger1}. 

A high-power, tunable laser could be used to measure inelastic scattering and map atmospheric transmission over the wavelengths of interest. Measurements at multiple wavelengths can determine relative atmospheric transmission without any need for information about the molecular density profile, which is usually taken from models such as the US Standard Atmosphere (COSEA 1976) for absolute single-wavelength measurements. We note that the transmission of the atmosphere at 1.064 microns is very high, and this is a natural reference point since high intensity Nd:YAG lasers emit at this wavelength.

A simultaneous multiband LIDAR system could detect both the elastic and inelastic return signals. Using a tunable laser would allow us to measure on and off band regions near the water absorption feature, thereby measuring the strength of that feature. The LIDAR system can be aligned with the main survey system, and can take data during the readout and slew of the survey instrument.

The ability to map out the vertical profile of attenuation might provide a way to distinguish between attenuation processes that are spectrally coincident, such as aerosols (concentrated in the lower regions of the atmosphere) and ozone (concentrated in the lower stratosphere). Additionally, even thin clouds at high altitudes can be detected as an over-density of scattering particles. This added height dimension could contribute to atmospheric models discussed in the next section.

LIDAR is a mature technique that is commonly used in the atmospheric sciences, but has not yet been applied to the astronomical extinction problem. The use of high-power lasers at astronomical observatories does pose issues of safety as well as interference with photometric measurements at sites with multiple telescopes. However, adaptive optics systems already in place use similar devices which indicates that such problems are surmountable. We suspect that the narrowband laser light from
a nsec pulse tunable laser (with $\delta \lambda \sim $0.5 nm) is probably better
suited to measuring the aerosol scattering rather than water vapor absorption. A faster laser pulse would be correspondingly broader, of course.

\section{MODELING THE TRANSMISSION OF THE ATMOSPHERE: MODTRAN}

Sophisticated radiative transfer codes exist to model the transmission properties of the Earth's atmosphere.  
Here we discuss MODTRAN,  the \emph{MOD}erate spectral resolution atmospheric \emph{TRAN}smittance code 
developed by the US Air Force Research Lab \citep{2001SPIE.4381..455A}.   
MODTRAN produces a transmission function, from optical to infrared,  for arbitrary lines of sight through 
the atmosphere.  The code can be used with an input atmospheric profile, in full spherical refractive geometry, 
to decompose the integrated airmass into the altitude-dependent opacity along the line of sight.
The calculations are based upon spectroscopic band models and provide a maximum 
resolution of 2 cm$^{-1}$.  
The band model approach is fast
but is limited to local thermodynamic equilibrium (LTE) conditions.  Non-LTE effects may be neglected in the lower
atmosphere, below 50km, but may become important for species found in the upper atmosphere.
The molecular species modeled by MODTRAN are derived from the HITRAN database 
\citep{2005JQSRT..96..139R} and include H$_{2}$O, CO$_{2}$, O$_{3}$, N$_{2}$O, CO, CH$_{4}$,
HNO$_{3}$, NO, NO$_{2}$, SO$_{2}$, O$_{2}$, N$_{2}$, NH$_{3}$. As noted by
\cite{Bailey07} and \cite{Adelman96}, the HITRAN database is not complete and it may be useful
to augment the line list with updated laboratory or observational results.
Aerosol scattering is modeled in discrete layers through the atmosphere.  Generic troposphere models 
(desert, rural, urban, maritime) are available as well as water vapor contributions from clouds.
User defined data may be substituted to extend the models to particular conditions of interest. 
  
MODTRAN is also capable of producing an emission calculation to model sky glow.  However, it is restricted
to the thermal contributions.  The important night sky emission lines are produced deep in the non-LTE regime
hundreds of kilometers up in the atmosphere and are not modeled.

\subsection{Blending measurements with models}

In our preliminary study, we have adapted MODTRAN for the atmospheric conditions of Mauna Kea Observatories.
This procedure will be extended to PS1 on Haleakala, and in principle, can be used to produce an atmospheric
model appropriate for any observatory.  
The Mauna Kea Weather Center provides real-time temperature, pressure and humidity measurements from the ground, 
as well as daily balloon radiosonde data to map the altitude profile.  We have used these measured pressure and 
temperature profiles to generate an input atmosphere definition for MODTRAN.  The Weather Center also provides the water vapor profile
 which can be incorporated into the model atmosphere directly.  To derive the concentrations of other molecular species, 
 we simply scale the concentrations  in the generic US Standard Atmosphere by the pressure and temperature.  

The atmospheric transmission models hide hundreds of free parameters.  We have 
explored the minimum set necessary to fit the measured atmospheric features in measured stellar spectra
with MODTRAN models and find that, with appropriate model atmosphere profiles, we can fit observations
by scaling only the water vapor content.

\begin{figure}
\plotone{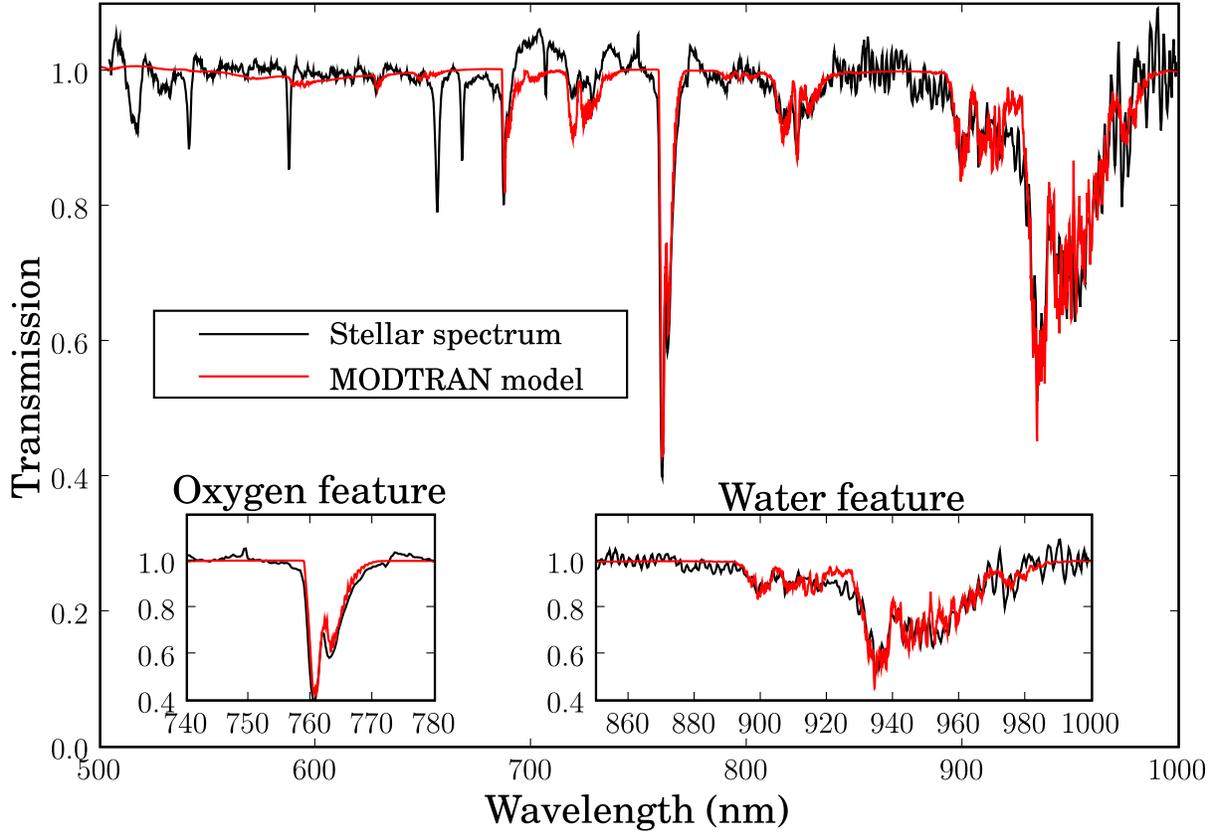}

\caption{This plot shows a spectrum we obtained on the UH 88 inch telescope, 
normalized to show absorption features. The inset panels show the comparison
between the observed spectrum and MODTRAN output, for the parameters
shown. We intend to pursue the idea of combining observations and 
models to determine the optical transmission function of the atmosphere.} 
\label{fig:spectrum}

\end{figure}

Figure~\ref{fig:spectrum} shows a measured O star spectrum from our Mauna Kea data set, overplotted with a MODTRAN model.
The spectrum has been flattened and normalized to isolate the line absorption features.
The primary absorption bands of interest in the optical are the oxygen line at 760 nm and the water band at
950 nm.  We tune the MODTRAN output to these features by first fixing the line of sight to match the observation,
and then scaling the oxygen and water vapor concentrations to match the absorption depths.
The water vapor is the dominant absorber in the MODTRAN model and we can put strong constraints
on it through the fit.   It is less sensitive to the oxygen concentration. Thus, through joint spectroscopic 
measurements and modeling, the temporal and spatial variations in water vapor can be effectively tracked.

Atmospheric transmission models can be integrated into an observatory using measurements from a dedicated 
instrument to monitor key absorption features.  As discussed, a simple system could consist of a dedicated imaging 
camera with narrow-band filters centered on strong atmospheric absorption bands.  The camera would
collect atmospheric transmission data in synchrony with the survey telescope operations, and provide
constraints on oxygen and water vapor concentrations as well as aerosol scattering.
These measurements could then be fit by an atmospheric modeling code, such as MODTRAN, to
produce a full, high-resolution transmission function representing a particular line-of-sight and time period.

\section{Summary and Next Steps}

This paper addresses the goal of achieving precise relative photometry over the course of
next-generation ``all-sky'' imaging surveys such as PanSTARRS and LSST.
Calibrations of absolute scales may also be possible, but the scientific goals for these
future surveys stress uniformity of the photometry across large scales on the sky, 
and the identification of time-dependent celestial
phenomena.

Our thesis is that significant improvement in photometric measurement,
perhaps even to the millimag level of relative precision, can be enabled by direct
measurement of atmospheric throughput. 
Several techniques for making such measurements are suggested, but all rely on identification
of a relatively small number of contributors to atmospheric extinction - the final four.
Each of these contributors leaves characteristic signatures in transmitted spectra.
It is the challenge of our approach to show how to detect and quantify these signatures
with sufficient temporal and spatial resolution to allow precise relative photometric
measurements to be made.
 
Molecular absorption is significant for only a few species
(specifically ozone, oxygen, OH, and water vapor) 
and occurs at well-known and characteristic wavelengths.
Water vapor column heights vary substantially with atmospheric conditions, but optical 
depths of remaining molecular constituents are simply given by barometric pressure.
The theory of Rayleigh-Carbannes molecular scattering is well developed, and the process
presents a strong and stable $\lambda^4$ signature that is easily recognized.
Measurement of aerosol scattering is considerably more troublesome.

We anticipate that the variability of both the aerosol content of the
atmosphere and the water content will present the dominant atmospheric 
limitation to precision photometry from next-generation surveys. This suggests that instrumentation development 
focus on these two concerns. 

Our analysis leads to some thoughts on future research.
Firstly, we conclude that, while it may be possible to ``tune-up" techniques and algorithms
with parasitic use of existing data, it will be necessary to carry out dedicated
measurements to test the ultimate capability of any of these to meet our goals.
This conclusion is not surprising since what we are trying to do has not been done before,
{\it i.e.} no existing data set has met the goals of these next-generation surveys.

In follow-on campaigns it will be most useful to carry out tests
with simultaneous spectroscopic
measurements of stars with photometric measurements of a number of standard targets.
While these measurements need not be carried out over the scales of future surveys,
they will need to be made with care and dedication if it is to be shown
that survey-wide millimagnitude photometry is possible.

\section{ACKNOWLEDGMENTS}

We thank Justin Albert, Tim Axelrod, James Battat, Yorke Brown, Kelly Chance, Chuck Claver,
Kem Cook, Doug Finkbeiner, Jim Gunn, Zjelko Ivezic, John McGraw, Gene Magnier, Eli Margalith, David Schlegel, 
Nick Suntzeff and Doug Welch for useful conversations and input. 
We are very grateful for the efforts of the AERONET 
consortium, and Brent Holben in particular, for both the establishment 
and operation of the 
aerosol monitoring site on Mauna Loa, and for making the AERONET data
readily available on the web. 
We have also used results from the Total Ozone Mapping Spectrometer (TOMS)
on the EarthProbe satellite, 
and we are grateful to the team that built and operated this instrument. 
We thank the LSST Corporation, Harvard University and the Department of 
Energy Office of Science and the US Air Force Office of Scientific Research for their support of this work.
The LSST design and development activity is supported
by the National Science Foundation under Scientific Program Order No. 9
(AST-0551161) through Cooperative Agreement AST-0132798. 
Additional support was provided through the NSF award to the
ESSENCE supernova cosmology project, through AST-0607485.
Portions of this
work were performed in part under Department of Energy contracts DE-AC02-
76SF00515, DE-AC02-98CH10886, DE-FG02-91ER40677 and W-7405-Eng-48.
Additional funding comes from private donations, in-kind support at Department
of Energy laboratories and other LSSTC Institutional Members.

\clearpage

\clearpage

\clearpage

\clearpage

\clearpage
\clearpage

\clearpage

\clearpage

\clearpage

\clearpage

\clearpage

\clearpage

\clearpage

\clearpage


\begin{thebibliography}


\bibitem[Adelman et al.(1996)]{Adelman96} Adelman, Saul J., Gulliver, Austin F. and Holmgren, David E. 
 \ 1996 in {\it Model Atmospheres and Spectrum Synthesis}, ASP Conference Series 108, 293.

\bibitem[Albert et al.(2006)] {Albert06} Albert, J et  al.~ \ 2006 astro-ph/0604339.

\bibitem[Anderson et al.(2001)]{2001SPIE.4381..455A} Anderson, G.~P., et al.~\ 2001, \spie, 4381, 455 

\bibitem[Anderson et al.(2003)]{Anderson03} Anderson, T.L. et al, \ 2003 Jour. of Atmos. Sci. 60, 119


\bibitem[Bailey, Simpson \& Crisp(2007)]{Bailey07} Bailey, J., Simpson, A. and Crisp, D. \ 2007
\pasp, 119, 228

\bibitem[BenZvi et al.(2007a)]{Auger1} BenZvi, S.~Y.~ et al. \ 2007 Nucl. Instrum. Meth., A574 171 . 

\bibitem[BenZvi et al.(2007b)]{Auger2} BenZvi, Segev et al. \ 2007 in 30$^{th}$ International Cosmic Ray Conference, and arXiv astro-ph/0706.1710

\bibitem[Bessell(1999)]{Bessell99} Bessell, Michael \ 1999 \pasp,  111, 1426

\bibitem[Burke et al.(2006)]{Burke06} Burke, David et al.~ \ 2006 \spie, 6267, 626715


\bibitem[COSEA(1976)]{COSEA} Committee on Extension to the Standard Atmosphere, \ 1976 ``US Standard
Atmosphere", US Govt Printing Office, Washington DC.

\bibitem[Dawsey et al.(2006)]{Dawsey06} Dawsey, M. et al.~\ 2006 \spie, 
6270 47. 

\bibitem[De Vaucouleurs \& Angione(1973)]{DeVauc73}De Vaucouleurs, G. and Angione, R.J. \ 1973 
\pasp, 86, 104

\bibitem[Eck et al.(1999)]{Eck99} Eck, T.F. et al.~\ 1999 Jour Geophy Res, 
103, 31. 
 
\bibitem[Everett \& Howell(2001)]{EH01} Everett, Mark E and Howell, Steve B., \ 2001 
\pasp, 113, 1428 
 
\bibitem[Frogel (1998)] {Frogel} Frogel, J. \ 1998 \pasp, 110, 200 


\bibitem[Hansen \& Travis(1974)]{Hansen} Hansen, James, E. and Travis, Larry D., \ 1974, Space Science Reviews, 16, 527.  

\bibitem[Hartman et al.(2005)] {Hartman05}  Hartman, J.D. et al.~\ 2005 \aj, 130, 2241 

\bibitem[Hadrava(2006)]{Hadrava06} Hadrava, P. \ 2006 A\&A 448, 1149

\bibitem[High(2007)]{High07} High, F.W. et al.~ \ 2007 (in prep).

\bibitem[Hinkle, Wallace and Livingston(2003)]{KPNO} Hinkle, K.H., Wallace, L., and 
Livingston, W., \ 2003 \baas, 203, 3803 
 
\bibitem[Holben et al.(1999)]{Holben99}  Holben, B.N., et al. \ 2001 , J. Geophys. Res., 106, 67
 
 
 \bibitem[Ivezic et al.(2007)]{Ivezic07} Ivezic, Z., et al.~ \ 2007 astro-ph/0703157, submitted to \aj.
 
 \bibitem[Jaross et al.(2003)]{Jaross03} Jaross, G. et al.~\ 2003 AGU Fall 
 Meeting abstract A21D-0993. 
 
 \bibitem[Kaiser et al.(2002)]{PS02} Kaiser, Nicholas et al.~\ 2002 \spie, 4836, 154 


 \bibitem[Krisciunas et al.(1987)] {KrisciunasMK} Krisciunas, K. et al.~\ 1987 \pasp, 99, 887
 
 \bibitem[Kurucz(1993)]{Kurucz} Kurucz, R., \ 1993, ATLAS9 Stellar Atmosphere Programs and 2 km/s grid. Kurucz CD-ROM No. 13. Cambridge, Mass.: Smithsonian Astrophysical Observatory.
 
 \bibitem[Kylling, Albold \& Seckmeyer(1997)]{Kylling97} Kylling, A., 
 Albold, A. and Seckmeyer, G., \ 1997, Geophys. Res. Lett 24, 397.
 
 \bibitem[Lantz et al.(2004)]{SNIFS04} Lantz, B et al.~\ 2004 \spie 5249, 146
 
 
\bibitem[LSST collaboration(2007)]{LSST} LSST collaboration, LSST Science Case.  {\tt http://www.lsst.org/Science/lsst\_baseline.shtml}
 
\bibitem[Magnier(2007)]{MagnierPS} Magnier, E, in  
The Future of Photometric, Spectrophotometric and Polarimetric Standardization, ASP Conference Series, Vol. 364, C. Sterken, ed, . San Francisco: Astronomical Society of the Pacific, 2007.

\bibitem[Measures(1984)]{Measures84} Measures, R. ''Laser Remote Sensing'', Wiley, 1984.  

\bibitem[Melfi(1972)]{Melfi72} Melfi, S.H.~ \ 1972 Applied Optics 11, No. 7  1605.
	  	  
\bibitem[Miknaitis et al.(2007)]{ESSENCE} Miknaitis, G. et al.~\ 2007 \apj,  in press, 
astro-ph/0701043


\bibitem[Padmanabhan et al.(2007)]{SDSS07} Padmanabhan, N. et al.~\ 2007 astro-ph/0703454

\bibitem[Pakstiene(2001)]{Pakstiene} Pakstiene, E. \ 2001 Baltic Astronomy, 10, 651


\bibitem[Rayleigh(1899)]{Rayleigh} Rayleigh, Lord, \ 1899 Phil. Mag. 47, 375Ð394

\bibitem[Reimann et al.(1992)] {Riemann92} Reimann, H.-G. et al.~ \ 1992 A\&A, 265, 360

\bibitem[Roosen \& Angione(1977)]{Roosen77} Roosen, Robert G. and Angione, Ronald J. \ 1977
\pasp, 89, 814

\bibitem[Rothman et al.(2005)]{2005JQSRT..96..139R} Rothman, L.~S., et al.~\ 2005, Journal of Quantitative Spectroscopy and Radiative Transfer, 96, 139 

\bibitem[Schuster, Parrao \& Guichard(2002)]{SanPedro} Schuster, W.J., Parrao, L and 
Guichard, J. \ 2002, Journal of Astronomical Data 8, 2. 

\bibitem[Sinyuk et al.(2007)]{Sinyuk07} Sinyuk, A.~ et al.~ \ 2007 Remote Sensing of Environment 107, 90

\bibitem[Stritzinger et al.(2005)]{Stritzinger05} Stritzinger, M et al.~ \ 2005 \pasp,  117, 810

\bibitem[Stubbs \& Tonry(2006)]{ST06} Stubbs, C. and Tonry, J. \ 2006 \apj,  646, 1436. 

\bibitem[Stubbs et al.(2007)]{CTIOlaser} Stubbs, C.W. et al.~ in
The Future of Photometric, Spectrophotometric and Polarimetric Standardization, ASP Conference Series, Vol. 364, C. Sterken, ed, . San Francisco: Astronomical Society of the Pacific, 2007., p.373
 
 \bibitem[Tonry et al.(2005)]{JT05} Tonry, John L. et al.~\ 2005 \pasp, 117, 281 
 
 \bibitem[Thomas-Osip et al.(2007)]{Campanas07} Thomas-Osip, J. et al.~ \ 2007 astro-ph/0706.2683 (accepted \pasp.)
 
 \bibitem[Tucker et al.(2006)]{SDSSpt} Tucker, D. et al.~ \ 2005 Astron. Nachr. 327, 821
 
 \bibitem[Vaughan et al.(1993)]{Vaughan93} Vaughan, G. et al.~ \ 1993 Applied Optics 32, 2758. 
 
 \bibitem[Wade \& Horne(1988)]{Wade88} Wade, Richard A. and Horne, Keith \ 1988 \apj, 324, 411
 
\bibitem[Young et al.(1991)] {Young} Young, Andrew T et al.~ \  1991 PASP 183, 221. 

\bibitem[Zimmer et al.(2006)]{ALE06}  Zimmer, P., et al.~ \ 2007  BAAS 209 1540.

\end{thebibliography}
\end{document}